\documentclass[10pt, conference, compsocconf]{IEEEtran}
\usepackage[utf8]{inputenc}
\usepackage{cite}
\usepackage{amsmath,amssymb,amsfonts}
\usepackage{algorithmic}
\usepackage{graphicx}
\usepackage{textcomp}
\usepackage{xcolor}

\usepackage{cite}
\usepackage{mdwmath}
\usepackage{mdwtab}
\usepackage{eqparbox}

\usepackage[tight, footnotesize]{subfigure}
\usepackage{fixltx2e}
\usepackage{stfloats}
\usepackage{url}
\usepackage{csquotes}
\usepackage{tikz}
\usepackage{tabu}
\usepackage{breakurl}

\usepackage{rotating} 
\usepackage{pdfpages}
\usepackage[hidelinks,
bookmarks=false,
pdftitle={Evaluating the End-User Experience of Private Browsing Mode},]{hyperref}
\usepackage{adjustbox}
\usepackage{color}
\usepackage{colortbl}
\usepackage[super]{nth}
\usepackage{array}
\usepackage{xspace}
\usepackage{booktabs}
\usepackage{tabularx}
\usepackage{makecell}
\usepackage{changepage}
\usepackage{tablefootnote}

\usepackage{amsmath,amssymb}
\usepackage{stackengine}
\usepackage{enumitem} 
\usepackage{lscape}
\usepackage{afterpage}
\usepackage{placeins}
\usepackage{amsmath,amssymb}
\usepackage{multicol}

\include{macros}

\newcommand{\eg}{e.g.}
\newcommand{\ie}{i.e.}
\newcommand{\etal}{et al. }
\newcommand{\num}{25\xspace}
\newcommand{\nn}{5\xspace}
\newcommand{\UiY}{1990s }

\newcommand{\IPBM}{30}
\newcommand{\point}[1]{\par\smallskip\noindent\textbf{#1. }}
\newcommand{\Cross}{$\mathbin{\tikz [x=1.4ex,y=1.4ex,line width=.2ex, red] \draw (0,0) -- (1,1) (0,1) -- (1,0);}$}
\newcommand{\Checkmark}{$\color{green}\checkmark$}
\newcolumntype{L}[1]{>{\raggedright\arraybackslash}p{#1}}

\renewcommand{\thefootnote}{\fnsymbol{footnote}}

\def\BibTeX{{\rm B\kern-.05em{\sc i\kern-.025em b}\kern-.08em
    T\kern-.1667em\lower.7ex\hbox{E}\kern-.125emX}}
\begin{document}

\title{Evaluating the End-User Experience of\\ Private Browsing Mode}

\author{\IEEEauthorblockN{Ruba Abu-Salma${}^{1,*}$, Benjamin Livshits${}^{2,3}$}\\
\IEEEauthorblockA{${}^1$~University College London (UCL)\\[0.25ex]${}^2$~Imperial College London\\[0.25ex]${}^3$~Brave Software}}

\maketitle
\renewcommand{\thefootnote}{\fnsymbol{footnote}}
\footnotetext{$^*$ The study was conducted while the author was an intern at Brave Software.}

\begin{abstract}
    Nowadays, all major web browsers have a private browsing mode. However, the mode's benefits and limitations are not particularly understood. Through the use of survey studies, prior work has found that most users are either unaware of private browsing or do not use it. Further, those who do use private browsing generally have misconceptions about what protection it provides.

However, prior work has not investigated \emph{why} users misunderstand the benefits and limitations of private browsing. In this work, we do so by designing and conducting a three-part study: (1) an analytical approach combining cognitive walkthrough and heuristic evaluation to inspect the user interface of private mode in different browsers; (2) a qualitative, interview-based study to explore users' mental models of private browsing and its security goals; (3) a participatory design study to investigate why existing browser disclosures, the in-browser explanations of private browsing mode, do not communicate the security goals of private browsing to users. Participants critiqued the browser disclosures of three web browsers: Brave, Firefox, and Google Chrome, and then designed new ones. We recruited~\num demographically-diverse participants for the second and third parts of the study.

We find that the user interface of private mode in different web browsers violates several well-established design guidelines and heuristics. Further, most participants had incorrect mental models of private browsing, influencing their understanding and usage of private mode. Additionally, we find that existing browser disclosures are not only vague, but also misleading. None of the three studied browser disclosures communicates or explains the primary security goal of private browsing. Drawing from the results of our user study, we extract a set of design recommendations that we encourage browser designers to validate, in order to design more effective and informative browser disclosures related to private mode. 
\end{abstract}

\section{Introduction} 
\label{Introduction} 

Prior work has extensively explored users' online privacy concerns when using the Internet~\cite{D1, D2, D3, D4, P1, D5, P2, D6}. For example, a survey of~1,002 US respondents (conducted by the Pew Research Center in 2013) found that respondents were concerned about their personal information being available online~\cite{P1}. Respondents also felt strongly about controlling who had access to their behavioural data and communications, including family members, partners, friends, employers, advertisers, and government agencies. In~2015, Angulo and Ortlieb conducted a user study to investigate users' concerns with regards to ``online privacy-related panic'' incidents~\cite{P2}. They identified~18 different incidents that would make participants panic or distress. Online tracking, reputation loss, and financial harm were the most frequently reported incidents by participants.

Prior work has also found that users are willing to take measures to protect their online privacy. In the same Pew Research Center survey~\cite{P1}, a clear majority~(86\%) of respondents reported they had taken steps to remove or hide their ``digital footprints,'' including clearing their browsing history and cookies. Further, Kang \etal conducted a user study to investigate how users would react to security and privacy risks~\cite{P3};~77\% of non-technical participants reported taking several measures to protect their ``digital traces,'' including the use of private browsing mode.

As we can see, users have serious concerns about their online privacy, and try to employ different strategies or use different privacy-enhancing tools to protect it. In this work, we focus on evaluating the end-user experience of one of these tools:~\textbf{private browsing mode}\footnote{~In this paper, we use the terms “private browsing mode,” “private browsing,” and “private mode” interchangeably.}. Private browsing is a privacy-enhancing technology (PET) that allows a user to browse the Internet without saving information about the websites they visited in private mode on their local device. 
As of today, all major web browsers have a private browsing mode.

Previous user studies have quantitatively -- mainly through survey studies -- investigated whether users are aware of private browsing, what they use it for, and whether they understand what protection it provides~\cite{PB0, PB1, PB2, PB3, PB4, PB5}. However, these studies have not investigated~\emph{why} most users misunderstand the benefits and limitations of private browsing mode. Further, the vast majority of recruited participants in these studies were unaware of or had not used private mode. In this work, we address these research gaps by designing and conducting a three-part study, where we recruited~\num demographically-diverse participants \textbf{(both users and non-users of private mode)} for the second and third parts of the study.

First, we use a hybrid analytical approach combining cognitive walkthrough and heuristic evaluation to inspect the user interface of private mode in different web browsers. \textbf{We identify several violations of well-known design guidelines and heuristics in the user interface of private mode.} We find some of these violations hampered the adoption and appropriate use of private mode.

Second, we conduct a qualitative, interview-based study to explore users' mental models of private browsing and its security goals. We find participants' conceptual understanding of the term ``private browsing'' influenced their mental models and usage of private mode in real life. Further, almost all participants did not understand the primary security goal of private browsing. Alarmingly, \textbf{we find that~\emph{all} participants who used private mode performed their private browsing activities while being authenticated to their personal online account (mainly their Google account to access certain online Google services)}, incorrectly believing their browsing or search history would be deleted after exiting private mode.

Third, we perform a participatory design study to investigate whether existing browser disclosures, the full-page explanations browsers present when users open a new private tab or window in private mode, communicate the security goals of private browsing to users. We ask participants to critique the browser disclosures of Brave, Firefox, and Google Chrome, and then design new ones. \textbf{We find that none of the three disclosures communicates the primary security goal of private browsing.} Our participants also pointed out that disclosures do not explain where information related to a private browsing session gets deleted from, and when.

\point{Contributions} Our primary contributions are:
\begin{itemize}
    \item We perform the~\emph{first} usability inspection of private mode in different web browsers using an analytical approach combining cognitive walkthrough and heuristic evaluation. We find the user interface of private mode violates several design guidelines and heuristics.
    \item We conduct the~\emph{first} qualitative user study to explore why most users misunderstand the benefits and limitations of private browsing. We do so by conducting an interview-based study with both users and non-users of private mode. We explore users’ mental models of private browsing and its security goals, and how these models influence users’ understanding and usage of private mode.
    \item We perform the~\emph{first} participatory design study to improve the design of browser disclosures related to private browsing mode. Prior work~\cite{PB1, PB4, PB5} has suggested that existing browser disclosures should be redesigned to better convey the actual benefits and limitations of private mode. In this paper, we do so by allowing our participants to take part in designing these disclosures; participants critiqued the browser disclosures of Brave, Firefox, and Google Chrome, explained why these disclosures are misleading, and then designed new ones.
    \item We extract a set of design recommendations that we encourage browser designers to validate (by implementing and testing), in order to design more effective browser disclosures.
\end{itemize}
\section{Related Work} \label{RelatedWork}
\subsection{User Studies of Private Browsing Mode} \label{UPBM} Prior work has~\textbf{quantitatively} (mainly through survey studies) investigated whether users are aware of private browsing, what they use it for, and whether they understand what protection it provides. In~\cite{PB1}, Gao~\etal conducted a survey of~200 Mechanical Turk (MTurk) respondents in the US, examining their private browsing habits. They found that one-third of respondents were not aware of private browsing. Those who had used private browsing reported using it for protecting personal information, online shopping, or visiting ``embarrassing websites.'' Further, most respondents had misconceptions about private browsing -- such as incorrectly believing that private mode protects from visited websites. Gao~\etal concluded that browsers do not effectively inform users of the benefits and limitations of private browsing, and that ``browser designers [should think of] various ways to [better] inform users.''

In~2017, DuckDuckGo, an Internet search engine, surveyed a sample of~5,710 US respondents, recruited via SurveyMonkey~\cite{PB2}. Respondents were asked to share their experience with private browsing. Again, one-third of respondents reported they had not heard of private browsing. Of those who had used private browsing, one-third used it frequently, and three-quarters were not able to accurately identify the benefits of private browsing. The report did not offer any recommendations beyond the study.

Using a similar study to~\cite{PB2}, Bursztein ran an online survey of~200 US respondents (via Google Consumer Surveys) in~2017~\cite{PB3}. He found about one-third of surveyed respondents did not know about private browsing. Of those who were aware of the technology, only~20\% had used it. Further, about one-half preferred not to disclose what they used private browsing for. Additionally, only~40\% claimed they used private browsing for its intended purpose: leaving no traces of the websites visited in private mode on the local machine. Bursztein concluded that the computer security and privacy community should raise awareness of what private browsing can and cannot achieve.

Recently, Wu~\etal surveyed~460 US respondents through MTurk~\cite{PB4}. Respondents were randomly assigned one of~13 different browser disclosures related to private mode. Based on the disclosure they saw, respondents were asked to answer a set of questions to assess their understanding of private mode. Wu~\etal found that existing browser disclosures do not sufficiently inform users of the benefits and limitations of private mode. They concluded that browser disclosures should be redesigned to better convey the actual protections of private browsing. They also argued that the term “private browsing” could be misleading. In this work, \textbf{we explore how users’ conceptual understanding of the term “private browsing" influences their understanding and usage of private mode in real life}.

Habib~\etal conducted a user study to observe the private browsing habits of over~450 US participants using software monitoring~\cite{PB5}. They then asked participants to answer a follow-up survey (using MTurk) to investigate discrepancies, if any, between observed and self-reported private browsing habits. They found that participants used private mode for online shopping and visiting adult websites. The primary use cases of private mode were consistent across observed and self-reported data. They also found that most participants overestimated the benefits of private mode, concluding by supporting “changes to private browsing disclosures.”


\point{Summary} Prior work has employed~\textbf{quantitative} methods – mainly through conducting surveys – to investigate whether users are aware of private browsing, what they use it for, and whether they understand what protection it provides (see Table~\ref{TRelatedWork} in Appendix~\ref{e}). However, prior work has not investigated~\textbf{why} users misunderstand the benefits and limitations of private browsing. Further, most recruited participants in prior user studies either were unaware of or had not used private mode. In this work, \textbf{we address these research gaps by designing and conducting a three-part user study: (1) the first usability inspection of private mode in different web browsers, (2) the first qualitative, interview-based user study, and (3) the first participatory design study. We also recruit both users and non-users of private mode.}%
\subsection{Mental Models} \label{MM} Users make computer security- and privacy-related decisions on a regular basis. These decisions are guided by users’ mental models of computer security and privacy. A mental model is someone’s understanding or representation of how something works~\cite{MM}. In their seminal paper, Saltzer and Schroeder provided eight principles that guide the design and implementation of computer security (or protection) mechanisms~\cite{MM1}. One of these principles is~\textit{psychological acceptability}: if there is a mismatch between a user’s mental image of a protection mechanism and how the mechanism works in the real world, the user will be unable to use the mechanism correctly. Wash and Rader proposed a new way to improve user security behaviour: instead of trying to teach non-technical users “correct” mental models, we should explore their existing models~\cite{MM2}. Wash conducted a qualitative study to investigate users’ mental models of home computer security~\cite{MM3}. He identified eight “folk models” of security threats that are applied by home computer users to make security-related decisions. Zeng~\etal qualitatively studied users’ security and privacy concerns with smart homes~\cite{MM4}. They found gaps in threat models, arising from limited technical understanding of smart homes.

Kang~\etal undertook a qualitative study to explore users’ mental models of the Internet~\cite{MM5}. Oates~\etal studied users’ mental models of privacy, asking end-users, privacy experts, and children to draw their models~\cite{MM6}. Through the use of interviews and surveys, Renaud~\etal investigated users’ mental models of encrypted email, and found that, in addition to poor usability, incomplete threat models, misaligned incentives, and lack of understanding of how email works are key barriers to adopting encrypted email~\cite{MM7}. Abu-Salma~\etal qualitatively and quantitatively explored users’ mental models of secure communication tools, and found that most users perceived encrypted communications as futile~\cite{MM8, MM9}. Wu and Zappala conducted a qualitative user study to investigate users’ perceptions of encryption and its role in their life~\cite{MM10}. They identified four users’ mental models of encryption that varied in complexity and detail. Krombholz~\etal qualitatively explored end-users and system administrators’ mental models of HTTPS, revealing a wide range of misconceptions~\cite{MM11}. Gallagher~\etal qualitatively studied experts and non-experts’ perceptions and usage of the Tor anonymity network, identifying gaps in understanding the underlying operation of Tor~\cite{MM12}.

\point{Summary} Prior work has explored users’ mental models of different computer security and privacy concepts and tools. In this work, \textbf{we qualitatively investigate users’ mental models of private browsing and its security goals}. We also give participants the option to~\textbf{draw their models}.
\subsection{Security and Privacy Design} \label{SPD} Within web browsers, prior work has investigated the design of alert messages and warnings~\cite{W1, W2, W3, W4, W5, W6, W7, W8}, browser security indicators~\cite{I1, I2, I3}, site trustworthiness~\cite{ST1, ST2}, privacy policies~\cite{PP1, PP2}, storage policies~\cite{SP}, and ad personalization~\cite{AP}.

However, prior work has heavily focused on the design of warning messages -- especially phishing warnings~\cite{W1, W2, W5, W6} and SSL warnings~\cite{W3, W4, W6, W7, W8} -- in order to capture users' attention, improve their comprehension, and warn them away from danger. For example, Egelman \etal recommended that phishing warning messages should be active (\ie interrupt the user flow) and should be distinguishable by severity~\cite{W2}. They also suggested it should be difficult for users to click-through phishing warnings, by requiring users to bypass several screens in an attempt to dissuade users from ignoring warnings. Additionally, Egelman and Schechter showed that changes to the look and feel of phishing warnings have resulted in more users noticing them~\cite{W5}. Felt~\etal recommended warning designers use opinionated design to improve user adherence to warnings~\cite{W8}.

Further, several researchers have focused on reducing user habituation to security warnings~\cite{H1, H2, H3}. Brustoloni and Villamarin-Salomon suggested the use of polymorphic and audited dialogues~\cite{HS1}. Bravo-Lillo~\etal explored the use of attractors~\cite{HS2}. Anderson \etal varied size, colour, and option order~\cite{HS3}.

\point{Summary} The aforementioned work has mainly focused on the design of browser warning messages to improve their efficacy. However, our study focuses on designing browser disclosures that sufficiently inform users of the benefits and limitations of a privacy-enhancing technology (private browsing). Although we draw inspiration from this work, \textbf{we answer a different important question of how to design browser disclosures to help users appropriately use private browsing mode}. We do so by employing~\textbf{participatory design~\cite{PD}}, asking participants to critique existing browser disclosures and design new ones. Unlike warning designers who have explored different ideas -- such as changing the design of a warning message or using attractors -- to improve user attention to and comprehension of warnings, \textbf{we choose, in this work, to engage users in the design of browser disclosures (related to private browsing mode)}.
\section{Private Browsing Mode} \label{PBM} Private browsing is a privacy-enhancing technology (PET) that allows the user to browse the Internet without \emph{locally} saving information (\eg, browsing history, cookies, temporary files) about the websites they visited in private mode~\cite{PB}. Nowadays, all major web browsers support private browsing. Different browsers refer to it using different names. For example, private browsing is known as~\textit{Incognito Browsing} in Google Chrome, \textit{InPrivate Browsing} in Microsoft Edge and Microsoft Explorer, and \textit{Private Browsing} in Brave, Firefox, Opera, and Safari. Further, Brave distinguishes between a~\textit{Private Tab} and a~\textit{Private Tab with Tor}, a new feature that was added in June~2018~\cite{BRAVE}.

\point{Private browsing goals} The primary security goal of private browsing is that a local attacker -- such as a family member, a friend, or a work colleague -- who takes control of the user’s machine \textit{after} the user exits a private browsing session should find no evidence of the websites the user visited in that session~\cite{PB}. That is, a local attacker who has (physical or remote) access to the user’s machine at time \textit{T} should learn nothing about the user's private browsing activities prior to time \textit{T}. Therefore, private browsing does not protect against a local attacker who controls the user's machine \textit{before} or \textit{during} a private browsing session; a motivated attacker (\eg, a suspicious wife) can install a key-logger or a spyware and monitor the user's (\eg, husband's) private browsing activities.

Further, private browsing does not aim to protect against a web attacker who, unlike a local attacker, does not control the user's machine but controls the websites visited by the user in private mode~\cite{PB}. Even if the user is not authenticated to an online service, a website can uniquely identify them through their client's IP address. Also, the user's various browser features -- such as screen resolution, timezone, and installed extensions -- can easily enable browser fingerprinting~\cite{PB} and, hence, website tracking.

Additionally, private browsing does not aim to hide the user’s private browsing activities from their browser vendor, Internet service provider (ISP), employer, or government.

To achieve the primary security goal of private browsing, once a user terminates a private browsing session, most web browsers claim to delete the user’s private browsing history, cookies, information entered in forms (\eg, login data, search items), and temporary files from the user’s local device. Further, some browsers do not locally store the bookmarks created and files downloaded in a private browsing session. Table~\ref{T0} summarizes the functionality of private mode in seven browsers.

\begin{table*}[htbp]
\caption{Private browsing functionality in recent web browser versions. A checkmark indicates an item is locally deleted once a user exits private mode, whereas a crossmark indicates an item is locally saved. \newline The table is not fully comprehensive; other items not shown include: browser cache, temporary files, HTML local storage, form auto-completion, client certificates, browser patches, phishing block list, and per-site zoom level. There has been no recent analysis of private browsing since the~2010 work of Aggarwal~\etal~\cite{PB}.}
\centering
\scriptsize
\setlength{\arrayrulewidth}{.2em}
\resizebox{\textwidth}{!}{
\begin{tabular}{lcccccccc}
\toprule
& \textbf{Brave}& \textbf{Firefox}& \textbf{Google Chrome}& \textbf{Internet Explorer}& \textbf{Microsoft Edge}& \textbf{Opera}& \textbf{Safari}\\
& \textbf{0.55}& \textbf{62.0.3}& \textbf{69.0.3497.100}& \textbf{11}& \textbf{44.17763.1.0}& \textbf{56.0.3051.36}& \textbf{12.0}\\
\midrule
\textbf{Browsing history}& \Checkmark& \Checkmark& \Checkmark& \Checkmark& \Checkmark& \Checkmark& \Checkmark\\
\textbf{Cookies}& \Checkmark& \Checkmark& \Checkmark& \Checkmark& \Checkmark &\Checkmark& \Checkmark\\
\textbf{Login data}& \Checkmark& \Checkmark& \Checkmark& \Checkmark& \Checkmark& \Checkmark& \Checkmark\\
\textbf{Search items}& \Checkmark& \Checkmark& \Checkmark& \Checkmark& \Checkmark& \Checkmark& \Checkmark\\
\textbf{Bookmarks}& \Cross& \Cross& \Cross& \Cross& \Cross& \Cross& \Cross\\
\textbf{Downloads}& \Cross& \Cross& \Cross& \Checkmark& \Checkmark& \Cross& \Checkmark\\
\bottomrule
\end{tabular}}
\label{T0}
\end{table*}

\point{Private browsing implementations} While all major web browsers have a private mode, each browser’s implementation of private browsing is different~\cite{PB}. Further, most browsers update their implementation based on user demand. For example, some browsers have recently added privacy features to help reduce website tracking (although protecting against website tracking is not a security goal of private mode). Brave has added onion routing (Tor) as an option to its private tabs~\cite{BRAVE}. Firefox disables third-party cookies to stop some types of tracking by advertisers~\cite{FIREFOX}. Opera also supports a VPN service~\cite{OPERA}.

Additionally, most implementations of private browsing are imperfect. Prior work in the field of computer forensics has found residual artifacts that remain on the user's local machine (after the user terminates their private browsing session) that could be used to identify the user’s private browsing activities~\cite{ATTACK1, Attack2, Attack3}. For example, Ohana and Shashidhar were able to recover all cached images, URL history, and usernames (with their associated accounts) from RAM and memory dumps for browsing activities performed in Internet Explorer's InPrivate mode (version 8.0)~\cite{ATTACK1}. 
For further attacks, we refer the reader to~\cite{PB}.

Although these attacks are crucial to consider in order to achieve overall browser security, they are not the focus of our work. In this paper, we evaluate the \textbf{end-user experience} of private mode.
\section{Methodology} \label{Methodology}

To explore why most users misunderstand the benefits and limitations of private browsing, we designed and conducted a three-part study:
\begin{enumerate}
    \item A hybrid analytical approach combining cognitive walkthrough and heuristic evaluation to inspect the user interface of private mode in different web browsers and identify any usability issues.
    \item A qualitative, interview-based user study to explore users’ mental models of private browsing and its security goals, and how these models influence users’ understanding and usage of private mode.
    \item A participatory design study to investigate why existing browser disclosures do not communicate the actual protection of private mode.
\end{enumerate}

For the second and third parts of the study, a trained researcher conducted all interviews in the UK in English between August~2018 and September~2018, by first conducting~\nn unstructured (open-ended) face-to-face interviews, lasting for~60 minutes on average each (see Table~\ref{T1} in Appendix~\ref{b}). The emerging themes from these~\nn interviews helped us design the study script we used to conduct our main interviews,~\num semi-structured face-to-face interviews lasting for~90 minutes on average each (see Table~\ref{T3} in Section~\ref{Dem}). When conducting the semi-structured interviews, the interviewer allowed participants to share their thoughts and ask any clarification questions. Further, the interviewer probed where appropriate, which is a common practice in semi-structured interviews~---~the interviewer uses a list of questions (\ie, a study script), but can ask follow-up questions as well as skip questions that have already been covered. Below, we describe our study script (see Section~\ref{P2} and Section~\ref{P3}).
\subsection{Research Questions} \label{RQ} In this paper, we answer the following research questions:
\begin{itemize}
    \item \textbf{RQ1:} Does private mode in different web browsers suffer from poor usability that hampers the widespread adoption and use of private browsing?
    \item \textbf{RQ2:} How do users perceive the term “private browsing?”
    \item \textbf{RQ3:} What are users’ mental models of private browsing (as a privacy-enhancing technology) and its security goals?
    \item \textbf{RQ4:} How do users perceive those who use private browsing? Do users perceive the routine use of private browsing as “paranoid” or “unnecessary?'' 
    \item \textbf{RQ5:} How do users’ mental models and perceptions influence their usage of private browsing?
    \item \textbf{RQ6:} Why do existing browser disclosures (related to private browsing) misinform users of the benefits and limitations of private browsing?
    \item \textbf{RQ7:} How can the design of browser disclosures be improved?
\end{itemize}
\subsection{Part~1: Identifying Usability Issues} \label{UI} Usability inspection has seen increasing use since the~\UiY as a way to evaluate the user interface of a computer system~\cite{HA1}. Usability inspection is aimed at finding usability problems in the user interface design and evaluating the overall usability of an entire system. Unlike empirical user studies (see parts~2 and~3 of our study below), a user interface is inspected by developers and evaluators without engaging users (\ie, without recruiting participants to assess the usability of a system). Evaluating a design with no users are present can identify problems that may not necessarily be revealed by an evaluation with users~\cite{HA1, HA4, HA5, HA6}. Although it is important to bring users into the design process, evaluating a design without users can also provide benefits.

There are several usability inspection methods. In this work, we use a hybrid approach combining cognitive walkthrough and heuristic evaluation to inspect the user interface of private mode in five different web browsers: Brave, Google Chrome, Microsoft Internet Explorer, Mozilla Firefox, and Safari. Both methods are actively used in human-computer interaction (HCI) research~\cite{HA2}.

\point{Cognitive Walkthrough} Cognitive walkthrough is a usability inspection method that focuses on evaluating a user interface design for its~\textbf{exploratory learnability}, a key aspect of usability testing~\cite{HA8} based on a cognitive model of learning and use~\cite{HA9, HA10}. First-time users of a system may prefer to learn how to use it by exploring it, rather than investing time in comprehensive formal training or reading long tutorials~\cite{HA12}. Cognitive walkthrough identifies problems that users could have as they approach an interface for the first time. It also identifies mismatches between how users and designers conceptualize a task, as well as how designers make assumptions about users’ knowledge of a specific task (which could, for example, impact the labelling of buttons and icons).

Cognitive walkthrough is task-specific, studying one or more user tasks. The process comprises a preparatory phase and an analysis phase. In the preparatory phase, evaluators decide and agree on the input to the cognitive walkthrough process: (1) a detailed description of the user interface, (2) the user interface’s likely user population and context of use, (3) a task scenario, and (4) a sequence of actions that users need to accurately perform to successfully complete the designated task. In the analysis phase, evaluators examine each of the actions needed to accomplish the task. The cognitive walkthrough process follows a structured series of questions, derived from the theory of exploratory learning, to evaluate each step (or action) in the workflow. A detailed overview of the cognitive walkthrough process can be found in~\cite{HA11}.

\point{Heuristic Evaluation} In 1990, Nielsen and Molich introduced a new method for evaluating a user interface, called heuristic evaluation~\cite{HA1}. Heuristic evaluation involves having usability evaluators judge dialogue elements in an interface against established usability principles (“heuristics”). Ten heuristics, derived by Nielsen and Molich, can be found in~\cite{HA1}. The use of a complete and detailed list of usability heuristics as a checklist is considered to add formalism. Jeffries \etal found that heuristic evaluation uncovered more issues than any other evaluation methods, whereas empirical user studies (see parts~2 and~3 below) revealed more severe, recurring, and global problems that are more likely to negatively affect the user experience of a system~\cite{HA17}.


\point{Hybrid Approach} To avoid biases inherent in either of the usability inspection methods, we used a hybrid approach combining two of the most actively used and researched methods: cognitive walkthrough and heuristic evaluation. Combining both task scenarios and heuristics was recommended by Nielsen~\cite{HA16} and Sears~\cite{HA18}. We describe the hybrid approach in Appendix~\ref{a}.
\subsection{Part~2: Exploring Mental Models and Usage} \label{P2} After inspecting the user interface of private mode and identifying several usability issues, we aimed to answer RQ2--RQ5 (see Section~\ref{RQ}), by qualitatively investigating participants' mental models of private browsing and its security goals, as well as exploring how participants perceived those who (regularly or occasionally) use private browsing. We also aimed to understand how participants’ mental models and perceptions influenced their understanding and usage of private mode.

Hence, we explored the following themes:
\point{Mental models of “private browsing”} We asked participants whether they have heard of the term “private browsing,” and, if so, whether or not they felt confident explaining what it meant. We then asked them to explain what it meant to browse privately. We provided participants with a large pad of paper and a~24-colour pack of markers, giving them the option to draw their mental models of private browsing. Further, we asked participants to describe the benefits and drawbacks, if any, of browsing privately.

By asking these questions, we aimed to investigate participants' conceptual understanding of the term ``private browsing,'' and how this understanding influenced their mental models and usage of private mode (as a privacy-enhancing technology), as we describe in detail next.

\point{Mental models of private mode (as a PET)} After exploring participants’ general mental models of the term “private browsing,'' we asked participants whether they had browsed in private mode and, if so, whether they felt confident explaining what it meant to open a private tab or window. We then asked them to explain the difference, if any, between default (non-private) browsing mode and private browsing mode.

We also aimed to understand how participants perceived the security goals of private mode. Hence, we asked them about the entities, if any, that could learn about their private browsing activities (\eg, visited websites in private mode), and how. We wanted to explore whether participants understood the primary security goal of private browsing: protecting against a local attacker who takes control of a user's machine after the user exits private browsing (see Section~\ref{PBM}).

\point{Perceptions of users of private mode} We then asked participants to explain how they perceived those who use, or would be interested in using, private mode. We aimed to investigate whether participants perceived the use of private mode as paranoid or unnecessary.

\point{Expectations} We asked participants to describe what they would expect from private mode. We also investigated whether participants' familiarity with private mode affected the robustness of their mental models. Therefore, we asked participants to list the web browsers they regularly used (as well as those they did not necessarily use) and that they considered having a private mode that met their expectations.

\point{Private browsing usage} Finally, we aimed to explore how participants' mental models and perceptions influenced their usage of private mode. Hence, we asked participants who used, or had used in the past, private mode to share their private browsing habits. We asked them what they used private mode for, how often they used it, and where they used it. We also asked them to explain what they liked and disliked about private mode.
\subsection{Part~3: Designing Better Browser Disclosures} \label{P3} After exploring our participants’ mental models and usage of private mode, we aimed to investigate why browser disclosures (related to private browsing) do not communicate the actual benefits and limitations of private browsing. We also sought to improve the design of existing browser disclosures. Hence, we performed a participatory design study to solicit new disclosure designs from our participants.

\point{Assessing participants’ knowledge of private mode (before tutorial)} To answer RQ6 and RQ7 (see Section~\ref{RQ}), we first asked our participants to take a short quiz to further test their knowledge of private browsing. We asked them to answer the following questions about a private browsing mode that works properly:
\begin{itemize}
    \item Private mode hides my browsing activities from [browser vendor].
    \item If I visited a website in private mode, the website would not be able to determine whether I was browsing in private or public mode.
    \item After I exited private mode, a family member would not be able to learn about my activities in private mode.
    \item Before I start browsing in private mode, a family member will not be able to learn about the websites I plan to visit in private mode.
    \item Private mode encrypts information I send and receive while browsing in private mode.
    \item Private mode hides my browsing activities from my school or employer.
    \item Private mode hides my identity from websites I visit.
\end{itemize}

We also asked participants whether they were familiar with the following items that appear on almost all of today’s browser disclosures, and whether they felt confident explaining what each item meant: browsing history file, cookies, search items, bookmarks, downloads, and temporary files.

\point{Giving a tutorial} We then gave participants a~15-minute tutorial, explaining the primary security goal of private browsing, the difference between default browsing mode and private browsing mode, and why private browsing does not protect against website fingerprinting and, hence, website tracking and ad targeting. Further, we explained the different items/files that most web browsers claim to delete once a user exits private mode (see Section~\ref{PBM}). We also explained the different privacy features that have been recently added by some web browsers (\eg, Brave's Private Tabs with Tor). Finally, we explained the difference between a private tab, a private window, and a private session.

\point{Assessing participants’ knowledge of private mode (after tutorial)} To evaluate whether participants’ knowledge of private browsing had improved after the tutorial, we asked participants to take the same quiz we gave them previously. However, we shuffled the questions to minimize bias.

\point{Critiquing existing disclosures} We then asked participants to critique existing browser disclosures (using the knowledge they acquired from the tutorial). We sought to get feedback on three disclosures, as well as solicit new disclosure designs from participants. Hence, we asked each participant to critique the browser disclosures of three web browsers: Brave, Firefox, and Google Chrome. 
To minimize bias, disclosures were assigned to each participant randomly. We chose these three disclosures because Firefox and Chrome were the most frequently-used browsers by our participants. Further, Brave was launched with privacy as a key selling point. 

We showed participants one disclosure at a time. We then asked them to describe what they felt about the disclosure, how useful they felt the explanation was, what about the explanation would make them decide to use or not use private mode, and what else they would like the disclosure to tell them or elaborate on. We gave participants green and red markers to highlight what they liked and disliked about the disclosure. We then showed participants the second disclosure and followed-up by asking the same questions we asked about the first disclosure they saw. We also asked participants to compare the second disclosure to the first one, and then explain whether they would be more or less likely to use private mode if they saw this disclosure or the prior one. Additionally, we showed participants the third disclosure and asked them the same questions we previously asked.

\point{Soliciting new disclosure designs} Finally, we performed a participatory design study to solicit new disclosure designs from our participants. We asked participants to describe private browsing as if they were explaining it to someone new to this privacy-enhancing technology. We prompted our participants as follows: ``We would like you to design a browser disclosure that clearly explains the benefits and limitations of private browsing. While designing, think about what would make you use private mode, what information you would want to know, what information you would want to omit, and how you would want the disclosure to look.'' We gave participants a large pad of paper and a~24-colour pack of markers to design their disclosures, giving them the option to draw.

We also asked participants to share their thoughts on the following names: ``Private Browsing,'' ``InPrivate Browsing,'' and ``Incognito Browsing,'' and suggest a new name, if any. 
\subsection{Recruitment} \label{Recruitment} In this work, our focus is to understand how \textbf{mainstream users} perceive private browsing and its security goals. This understanding is crucial to design browser disclosures that sufficiently inform the general public of the benefits and limitations of private browsing. We do not investigate how a specific at-risk user group -- such as activists, journalists, or whistle-blowers -- perceive and use private browsing. However, we have documented our study protocol step-by-step, meaning that it can be replicated with different user groups in varying contexts.

To recruit our participants \textbf{(for the second and third parts of the study\footnote{~We did not recruit participants for the first part of the study (usability inspection).})}, we posted flyers and distributed leaflets in London (UK). We asked interested participants to complete an online screening questionnaire, which about~500 completed. We aimed to recruit a demographically-diverse sample of participants. Hence, we included a number of demographic questions about gender, age, race, educational level, and employment status. We also assessed participants' technical knowledge; we considered participants as technical if two out of three of the following were true~\cite{R1}: (1) participants had an education in, and/or worked in, the field of computer science, computer engineering, or IT; (2) they were familiar with or an expert in at least one programming language (\eg, C++); (3) people usually asked them for computer-related advice. Further, we provided participants with a list of different web browsers, and then asked which browsers they used, what they used each browser for (in case they used multiple browsers), which browser they used the most, and how many hours they spent daily on their desktop and mobile phone browsing.

Additionally, we asked participants to list the digital security requirements they had at school or work, how often they received cybersecurity training, and whether they felt at risk due to their school work or job duties. In~\cite{J8}, Gaw \etal found that people perceived the ``universal, routine use of encryption as paranoid.'' In this work, we aimed to explore whether our participants perceived the everyday use of private mode as paranoid and unnecessary.

We first conducted and analyzed~\nn unstructured interviews (to help us design the study script, which we describe in detail in Section~\ref{P2} and Section~\ref{P3}), followed by~\num semi-structured interviews (our study's main interviews). 
\subsection{Pilot Study} \label{PILOT}

\point{Quiz piloting} After developing an initial questionnaire of our quiz (see Section~\ref{P3}), we conducted interviews with~\nn demographically-diverse participants (see Table~\ref{T2} in Appendix~\ref{c}). Cognitive interviewing is a method used to pre-test questionnaires to glean insights into how participants might interpret and answer questions~\cite{PS1}. After answering each quiz question, participants were asked to share their thoughts and answer the following: ``Was this question difficult to understand or answer?;'' ``How did answering the question make you feel?'' We then used the findings to revise our quiz, and evaluate question wording and bias.

\point{Main study piloting} To pre-test the second and third parts of our study (pre-screening questionnaire, study script, and quiz), we conducted a small-scale pilot study of~\nn semi-structured interviews. We used the common practice of convenience sampling~\cite{PS1}, by selecting~\nn colleagues for the pilot study. Additionally, we asked~10 computer security and privacy researchers and experts to review the study. We used the findings to identify potential problems (\eg time, cost, adverse events) in advance prior to conducting the full-scale study.

Drawing from the findings of our pilot study, we made the following study design changes:
\begin{itemize}
    \item We decided to give participants a~10-minute break between the second (interviews) and third (participatory design) parts of the study, to reduce interviewee fatigue and inattention~\cite{DS}.
    \item As part of the participatory design study, we asked participants to take a quiz (before our tutorial) to assess their knowledge of private mode. Based on the pilot study findings, we decided to give participants the same quiz~\textit{after} the tutorial, to assess whether or not participants’ knowledge had improved before they started analyzing and critiquing browser disclosures.
    \item We first aimed to ask participants to critique the browser disclosures of five web browsers: Brave, Google Chrome, Microsoft Internet Explorer, Mozilla Firefox, and Safari (as part of the participatory design study). However, due to interviewee fatigue (as per our pilot study findings), we decided to analyze the disclosures of three browsers – Brave, Chrome, and Firefox – based on how popular the browser is and how it advertises itself (\eg, as fast, safe, or private).
\end{itemize}
\subsection{Data Analysis} \label{DataA}

\point{Part~1 of study} Two researchers inspected the user interface of private mode in Brave, Google Chrome, Microsoft Internet Explorer, Mozilla Firefox, and Safari. They did so independently before discussing the findings and aggregating all the uncovered issues in a larger set. 

\point{Parts~2 and~3 of study} To develop depth in our exploratory research, we conducted multiple rounds of interviews, punctuated with periods of analysis and tentative conclusions~\cite{GT1}. In total, we conducted, transcribed (using an external transcription service) and analyzed all~\nn unstructured and~\num semi-structured interviews (the study’s main interviews). We observed data saturation~\cite{DS, GT5} between the~\nth{20} and the~\nth{25} semi-structured interview; \ie, no new themes emerged in interviews~20--25, and, hence, we stopped recruiting participants. Data saturation has attained widespread acceptance as a methodological principle in qualitative research. It is commonly taken to indicate, on the basis of the data that has been collected and analyzed, further data collection and analysis are unnecessary.

Two researchers independently coded all interview transcripts and image data using grounded theory~\cite{GT1}, an open-ended method to discover explanations, grounded in empirical data, about how things work. The researchers created two codebooks: one for the interview transcripts and one for the image data. After creating the final codebook, they tested for the inter-rater reliability (or inter-coder agreement). The average Cohen's kappa coefficient ($\kappa$) for all themes in the interview transcripts and image data was~0.77 and~0.89, respectively. A $\kappa$ value above~0.75 is considered excellent agreement~\cite{KAPPA1}. 
\subsection{Ethics} Our study was reviewed and approved by our organization’s ethics committee. Before each interview, we asked participants to read an information sheet that explained the high-level purpose of the study and outlined our data-protection practices. We also asked participants to sign a consent form that presented all the information required in Article~14 of the EU General Data Protection Regulation (GDPR). Participants had the option to withdraw at any point during the study without providing an explanation. We paid each participant~\pounds\IPBM.
\section{Results} \label{R} In this section, we present the results of our study. We first describe the demographics of participants recruited for the second and third parts of our study (Section~\ref{Dem}). We then discuss the results of each part of our three-part study (Sections~\ref{R1}, \ref{R2}, and \ref{R3}).
\subsection{Demographics} \label{Dem} Table~\ref{T3} summarizes the demographics of our sample (n=\num participants). We interviewed~10 male,~13 female, and two non-binary participants. Participants’ ages ranged from~18 to~75.~12 identified as white, four as black, four as Asian, three as Hispanic, and two as mixed-race.~11 reported having a college (or an undergraduate) degree, and eight a graduate (or postgraduate) degree. Two reported having secondary (or high-school) education, and three some post-secondary education (\ie, some college education without a degree). One participant mentioned having vocational training (VOC). Nine participants were either high-school or university students,~12 employed, two unemployed, and one retired. One participant preferred not to indicate their employment status. According to the definition we used to assess our participants’ technical knowledge (see Section~\ref{Recruitment}),~17 qualified as technical.

Our participants used a wide range of web browsers (both on desktop/laptop and mobile phone). Google Chrome was the most used browser by participants, followed by Safari, Mozilla Firefox, Microsoft Internet Explorer, and Brave, respectively. Three participants (P01; P03; P25) used the Tor browser. We noticed younger participants used (or had used in the past) multiple web browsers, whereas older or less-educated participants often used one browser – mainly Safari due to its compatibility with Apple devices.

Participants daily spent between five and~17 hours (mean=11.70 hours) browsing the Internet. Desktop/laptop browsing overtook smartphone surfing, with the exception of three participants (P02; P12; P16). Further, most participants (22 out of~\num) used multiple browsers for various reasons. For example,~13 reported they used one browser for social activities and used a different one for work-related activities.

Prior user studies (see Section~\ref{UPBM}) have aimed to understand what people use private mode for. However, the vast majority of participants recruited for these studies were unaware of or had not used private mode. In our work, we recruited and interviewed both users and non-users of private mode.~19 participants reported they used (or had used in the past) private mode. Three (P12; P16; P24) were aware of private mode, but had not browsed in it. Three (P02; P11; P23) did not know private mode existed.

Finally, we note P01, P03, P18, and P25 identified as computer security and privacy experts. Hence, they did not necessarily represent mainstream users.
\begin{table}[htbp]
\caption{Semi-Structured Interview Participant Demographics}
    \scriptsize
    \centering
    \setlength{\arrayrulewidth}{.2em}
    \setlength{\tabcolsep}{4pt}

    \begin{tabular}{llllll}
    \toprule
    &\textbf{Gender}&\textbf{Age}&\textbf{Race}&\textbf{Education}&\textbf{Employment}\\ \midrule
    \textbf{P01}&Male& 25--34& White& Ph.D.& Student\\
    \textbf{P02}&Male& 45--54& Mixed race& B.A.& Unemployed\\
    \textbf{P03}&Male& 45--54& White& Ph.D.& Unemployed\\
    \textbf{P04}&Female& 18--24& Black& High-school& Student\\
    \textbf{P05}&Female& 25--34& White& B.A.& Employed\\
    \textbf{P06}&Male& 35--44& White& M.Sc.& Employed\\
    \textbf{P07}&Female& 18--24& White& B.A.& Employed\\
    \textbf{P08}&Female& 25--34& Asian& High-school& Student\\
    \textbf{P09}&Male& 18--24& Asian& M.Sc.& Employed\\
    \textbf{P10}&Male& 25--34& White& Some college& Employed\\
    \textbf{P11}&Female& 25--34& White& M.Sc.& Employed\\
    \textbf{P12}&Female& 45--54& White& Some college& Employed\\
    \textbf{P13}&Male& 25--34& Mixed race& B.A.& Employed\\
    \textbf{P14}&Male& 18--24& Hispanic& B.A.& Employed\\
    \textbf{P15}&Female& 25--34& Asian& B.Sc.& Other\\
    \textbf{P16}&Female& 45--54& Black& VOC& Employed\\
    \textbf{P17}&Female& 18--24& White& Ph.D.& Student\\
    \textbf{P18}&Non-binary& 35--44& White& M.Sc.& Employed\\
    \textbf{P19}&Female& 35--44& Black& B.Sc.& Self-employed\\
    \textbf{P20}&Male& 18--24& White& Some college& Retired\\
    \textbf{P21}&Male& 25--34& White& VOC& Student\\
    \textbf{P22}&Male& 18--24& Asian& Ph.D.& Student\\
    \textbf{P23}&Female& 25--34& White& M.Sc.& Student\\
    \textbf{P24}&Female& 25--34& Black& B.Sc.& Student\\
    \textbf{P25}&Female& 25--34& Hispanic& Some college& Student\\
    \bottomrule
    \end{tabular}
    \label{T3}
\end{table}
\subsection{Part~1: Identifying Usability Issues} \label{R1}

We used an analytical approach combining cognitive walkthrough and heuristic evaluation to inspect the user interface of private mode in five different web browsers (desktop versions): Brave, Google Chrome, Microsoft Internet Explorer, Mozilla Firefox, and Safari. Our findings are as follows:

\point{Public mode as the default mode} In all modern web browsers (including the ones we inspected), the default mode is the public one. To browse in private mode, users need to select (from a hidden drop-down list) “New Incognito Window” in Brave and Google Chrome, or “New Private Window” in Microsoft Internet Explorer, Mozilla Firefox, and Safari. We hypothesize (and find in Section~\ref{R2}) that most users are unaware of the hidden drop-list, which explains why most users do not know about private mode. This violates Nielsen's heuristic of~\textit{visibility of system status}~\cite{HA2} and \textit{aesthetic and minimalist design}~\cite{HA2}. %

\point{Multiple windows and tabs} Users cannot open a private tab in a public window, and vice-versa; that is, users can only open public (private) tabs in public (private) windows – which we regard as good user interface design. Further, users can only re-open the most recently-closed public tabs, and not private ones.

Although users can open multiple public and private windows, feedback is minimal. For example, in Safari, when users enter private mode, there is no appropriate feedback -- through the user interface -- that communicates to users that they are currently browsing in private mode. There is only a short line of text (using a small font size) at the top of the page that says: “Private Browsing Enabled," violating Nielsen's heuristic of~\textit{visibility of system status}~\cite{HA2}. In Brave and Mozilla Firefox, the background changes from white to purple. Both browsers do not explain why the color purple was chosen by browser designers.


\point{Use of jargon} Both Brave and Google Chrome refer to private mode as “Incognito window,” and Microsoft Internet Explorer, Mozilla Firefox, and Safari as “private window.” This violates Nielsen's heuristic of~\textit{match between the system and the real world}~\cite{HA2}, making the assumption that users' understanding and interpretation of words would be the same as browser designers and developers. We also hypothesize that users would build their own mental models of private mode when encountering these terms, which could strongly impact how they would perceive and use private mode in real life. We explore these models in depth in~\ref{R2} and~\ref{R3}.

\point{Wordy browser disclosures} When users enter private mode, a browser disclosure is shown to users. The disclosure is meant to explain the benefits and limitations of private browsing. However, the disclosures of all inspected browsers (except that of Firefox) are lengthy and full of jargon, violating Nielsens' heuristic of~\textit{match between the system and the real world}~\cite{HA2}. Further, browser disclosures do not explain the primary security goal of private mode. In Firefox, the disclosure is relatively short, but, also, does not explain the security goal of private mode.

Further, in all five browsers, users are presented with these disclosures only once (when they open a private window or tab), violating Nielsen's heuristics of~\textit{recognition rather than recall}~\cite{HA2} and \textit{help and documentation}~\cite{HA2}.

In Section~\ref{R3}, we present the results of our participants who critiqued existing browser disclosures and suggested several design options for improvement, as we explain later in the paper.

\point{Private browsing and Tor} Brave has recently added Tor to its private windows. Brave users can now open a “New Window,” “New Incognito Window,” or “New Private Window with Tor.” Both Incognito windows and private windows with Tor have the same purple background and lengthy disclosures, which could lead users to browse in one instead of the other, violating Nielsen's heuristic of~\textit{visibility of system status}~\cite{HA2}. Further, the browser disclosures of both windows do not clearly explain how private mode and Tor are two different privacy-enhancing technologies.





\subsection{Part~2: Exploring Mental Models and Usage} \label{R2} The main purpose of qualitative research is to explore a phenomenon in depth, and not to investigate whether or not findings are statistically significant or due to chance~\cite{PS1}. Although we report how many participants mentioned each finding as an indication of prevalence, our findings are not quantitative. Further, a participant failing to mention a particular finding does not imply they disagreed with that finding; they might have failed to mention it due to, for example, recall bias~\cite{PS1}. Thus, as with all qualitative data, our findings are not necessarily generalizable beyond our sample. However, they suggest several future research avenues, and can be later supplemented by quantitative data.

In this section and the next section (Section~\ref{R3}), we present the results of the second and third parts of the study (n=\num participants).

\point{Mental models of ``private browsing''} We aimed to investigate our participants' conceptual understanding of the term ``private browsing.''~18 out of~\num (a clear majority) had heard of the term, and~17 felt confident explaining what the term meant\footnote{~It is worth to mention that only three out of the 17 confident users associated the term ``private browsing'' with private mode. We speculate this is because these three participants used private mode frequently.}.~16 out of~17 were users of (or had used in the past) private mode. One participant (P11) was a non-user.

We then asked all participants to explain what ``private browsing'' meant to them.~5 out of~\num associated the term with private browsing mode, mentioning the following: ``the window that has a man with a coat and a pair of eye glasses'' (x4); ``going undercover or incognito'' (P04). All five participants were referring to the ``Incognito Window'' in Google Chrome. Further, five participants thought of the term in connection with network-encrypted communications or secure browser connections (\ie webpages running HTTPs), three with end-to-end encrypted communications, three with anonymous communications (using Tor or VPN), and three with user authentication (both one-factor and two-factor authentication). One participant (P17) associated ``private browsing'' with both network encryption and authentication. Additionally, P15 described the term as the ability to browse the Internet ``without getting infected with a virus.''

Further, eight participants mentioned the terms ``privacy'' and ``online privacy'' to explain what ``private browsing'' meant to them: P01--P05, P07, and P12--P14 defined the term as having control over how users' online information is handled and shared with others. P09, P20, P22, and P24 referred to the term as the ability to manage and ``regulate'' one's social space.

The drawings in Appendix~\ref{e} explain some of our participants' mental models of ``private browsing.''

We below show how participants' mental models of ``private browsing'' influenced their understanding and usage of private mode in real life. 

\point{Mental models and usage of private mode (as a PET)} \label{R2} After exploring our participants' conceptual understanding of the term ``private browsing,'' we aimed to investigate how this understanding influenced participants' mental models and usage of private mode (as a privacy tool). We identified three types of users: regular users, occasional users, and former users. We explain each type as follows: 

\textit{1. Regular users:} Two participants (P01 and P17) were regular users of private mode. They performed all their browsing activities in private mode. They described themselves as ``paranoid'' and ``cautious.'' P01 mentioned that the routine use of private mode made them feel ``safer'' and ``more comfortable.'' Further, P01 used Safari's private mode to protect against shoulder-surfing. They explained that Safari does not have a visual user interface element that indicates a user is currently browsing privately. However, when probed, P01 (as well as P17) did not know that staying in private mode for a long duration of time can easily enable fingerprinting and, hence, website tracking (a threat that both participants thought they were protected against by regularly browsing in private mode). 

\textit{2. Occasional users:} Out of~\num, 15 participants used private mode occasionally depending on their browsing activities and the websites they visited. They did not necessarily use the mode to visit ``embarrassing websites.'' Many used private mode for online shopping (\eg, purchasing a surprise gift for a family member or a friend), logging into an online service using a different account, and/or debugging software. 

\textit{3. Former users:} Two participants (P13 and P19) reported they had used private mode before, but stopped using it for the following reasons:
\begin{itemize}
    \item \textit{Lack of utility.}
P13 stopped using private mode because they thought that web browsers did not allow extensions to run in private mode (although users can manually enable extensions in private mode in most browsers).
    \item \textit{Lack of usability.}
P13 and P19 mentioned that entries added to the history file would get deleted if they exited private mode, negatively impacting user experience. P13 also mentioned that private mode is ``useless'' because users could delete information about websites visited in default mode by manually clearing their browsing history file and cookies (a view shared by P12 and P16). 
    \item \textit{Misconceptions about private mode.}
P13 perceived those who used private mode as people who ``had something to hide'' or ``were up to no good,'' influencing P13's decision to stop using private mode because they did not want to be perceived by others in their community as ``a cybercriminal'' or ``a terrorist.'' Many participants shared this perception, as we discuss later in this section.
\end{itemize}

Several participants (17 out of~\num) reported they mainly used private mode in public spaces, mainly coffee shops, libraries, and airports. They also performed browsing activities they regarded as sensitive in private mode. For example,

``\textit{I usually use Incognito in \dots you know \dots in Google when I work at [coffee shop] because I connect to the Internet using insecure or public Wi-Fi. My laptop consistently warns me. So, I use Incognito to encrypt my data and hide it from people around me \dots Better to be safe!}'' (P05)

``\textit{I usually use the public or \dots shared workstations in my school's library. You don't need to login because there is one account shared by all students. I usually open a private tab or \dots window -- I don't know -- to download files that I want to be removed after I close the browser \dots By the way, I also use a private window to send an encrypted email.}'' (P17)

P17 is a regular user of Safari that locally deletes files downloaded in its private mode. However, P17 did not notice he was using Firefox on the library's computer, which does not delete private browsing downloads.

``\textit{I usually make a bank transfer or access my personal online accounts -- you know, like Facebook -- when I use one of the computers that all passengers can use \dots I am talking about the computers you find in an airport lounge \dots I open a private window.}'' (P07)

``\textit{I use Incognito to search for new jobs. As you know, I do not want my boss or company to know \dots}'' (P18)

``\textit{If I do not have Tor installed, I will use Incognito.}'' (P09)

We also found six participants who tended to use private mode to visit malicious webpages. For example,

``\textit{I sometimes encounter a message that warns me from accessing a bad webpage. I usually ignore the warning and open the page in a private window \dots Feels safer!}'' (P14)

Alarmingly, \textbf{we found that \emph{all} participants who identified as either regular or occasional users of private mode (total=17 participants) performed their private browsing activities while being authenticated to their personal online account (\eg, their Google or YouTube account)}, believing their search history would be deleted after exiting private mode). 

Additionally, we found that some participants perceived those who use private mode as people who ``care about their online privacy,'' ``have something to hide'' (\eg, journalists, activists, dissidents), or ``are up to no good'' (\eg, cybercriminals, terrorists). These inappropriate mental models and misperceptions partially explain why most users overestimate the protection private mode offers.

\textbf{To summarize the findings above}, most participants found utility in private mode (\eg, online shopping, debugging software). However, our participants' conceptual understanding of the term ``private browsing'' negatively influenced their usage of private mode in real life. Many incorrectly believed that private mode could be used to send encrypted email, achieve online anonymity, or simply access a phishing webpage because it ``felt safer'' to do so. 

\point{Security goals of private mode} We aimed to further investigate how participants perceived the security goals of private mode. Thus, we asked participants about the entities, if any, that could learn about their private browsing activities, what they could learn, and how.

All, but three participants (P03; P18; P25) who identified as security/privacy experts, did not understand what private mode could and could not achieve (\ie, did not recognize the primary security goal of private browsing).

Many participants (19 out of~\num) believed that a family member, a partner/a spouse, a friend, or a work colleague would not be able to learn about the websites they visited in private mode ``whatsoever'' (P01). Ten mentioned that this would only be possible if the entity was ``technically-sophisticated.'' Only P03, P18, and P25 (as mentioned above) correctly explained that private mode protected against a local attacker \emph{after} the user exited private mode.

Several participants (12 out of~\num) believed that a browser vendor (\eg, Google, Microsoft) could not learn their private browsing activities, citing the following statement that appears on most browser disclosures: ``[Browser vendor] won't save your information \dots'' Further, seven participants believed that private mode would hide their browsing activities from the employer, six from the ISP, and six from intelligence services and governments.

As we can see, participants' perceptions partially explain why several participants perceived those who used private mode as paranoid or up to no good.

\point{Expectations} We then asked participants what they expected from private mode. Again,~19 expected that anyone who had access to their machine should find no evidence of the websites visited privately. Additionally,~10 expected that a private mode that worked properly would not link their browsing activities in private mode to those in public mode.~13 also expected that a private mode would protect them from all types of website tracking and ad targeting. Interestingly, five participants expected a website visited in private mode would not be able to determine whether the user is currently browsing privately or not. 

Although some browsers, such as Brave, have added privacy features to reduce online tracking, no browser meets all participants' expectations. However, we argue that participants' expectations were high because they overestimated the benefits of private mode. 
\subsection{Part~3: Designing Better Browser Disclosures} \label{R3} We aimed to investigate why existing browser disclosures do not communicate the actual benefits and limitations of private browsing. To further test participants' knowledge of private mode, we asked them to take a short quiz (see Section~\ref{Methodology}). Participants performed poorly with an average score of~3.21/7.00. Most participants (21 out of~\num) overestimated the benefits of private mode.

We also asked participants to explain the following items that appear on most browser disclosures: history file, cookies, and temporary files. We found that although all participants correctly described a browsing history file, most participants (21 out of~\num) either had not heard of a cookie or a temporary file, or did not feel confident explaining what these items meant (in the context of private browsing). These findings suggest that most participants did not understand the functionality of private browsing (see Section~\ref{PBM}), a finding recently echoed by~\cite{PB4}. \textbf{However, we argue (in Section~\ref{Discussion}) that users do not need to understand private browsing functionality in order to use private mode correctly}.

We then gave our participants a~15-minute tutorial, and asked them to take the same quiz again. Participants' quiz performance significantly improved (mean=~6.31/7.00), which was an indication that participants could use the knowledge they newly acquired to critique existing browser disclosures (related to private browsing) and then design new ones, as we discuss next. 

Hence, we asked participants to critique the disclosures of Brave, Firefox, and Google Chrome. We describe their views below:

\point{Private mode} Most participants (20 out of~\num) criticized Firefox for describing their private mode as ``a private window.'' Further,~17 participants pointed out that although both Brave and Google Chrome name their private mode ``Incognito,'' they still use the phrase ``browse privately'' in the first sentence of its browser disclosure, which is ``misleading.''

Moreover,~19 participants were confused about when information (\eg, cookies, search items) about websites visited in private mode gets deleted: after ``closing a private tab?'' (P03), ``closing all tabs?'' (P09), ``closing a [private] window?'' (P11), ``closing a session?'' (P04; P11; P13; P21), or ``shutting down a browser?'' (P09; P14; P17; P20; P21; P22; P24). Also, five participants questioned whether or not one private session would be shared across multiple windows or tabs.

We also asked participants to suggest a new name for private mode, if any. All participants came up with random names: ``non-private,'' ``everything but private,'' ``insecure,'' ``random mode,'' and ``useless.'' Although all participants agreed that the term ``private browsing'' is misleading, there was no clear winner among the names they suggested.

\point{Primary security goal} The vast majority of participants (21 out of~\num) pointed out that none of the three disclosures explained the primary security goal of private browsing. Seven participants pointed out that although the Chrome disclosure says that ``[a user's] private browsing activity will be hidden from users sharing the same device,'' it does not explain that a user of the machine could easily monitor other users' activities by infecting the machine with a malware. 

Many participants (17 out of~\num) also mentioned that browser disclosures should mention all types of attackers that could violate the security policy of private browsing. They reported that all browser disclosures mention a subset of all possible attackers, and not the complete set.

\point{Private browsing functionality} Several participants (16 out of~\num) criticized the use of the following statement by all three disclosures: ``[vendor] will save/won't save the following information.'' Participants explained that the statement implied the vendor will not save information on its servers after exiting private mode. Yet, the true meaning of the statement is that the vendor will \textit{only} delete private browsing-related information from the user's local device, and not necessarily from the vendor's servers.

Further, about two-thirds of participants (17 out of~\num) suggested that the detailed technical explanation of private browsing functionality (\eg, whether cookies or temporary files are stored or not after exiting private mode) should be deferred until the primary security goal is explained in detail, which is none of the disclosures critiqued does. Participants mentioned that browser disclosures should explain (in bullet points) \textbf{what} protection private mode can and should offer (protecting from a local adversary). Yet, browser disclosures describe \textbf{how} this protection is achieved (\eg, by deleting cookies), without explaining \textbf{what} protection private mode offers.

\point{Tracking protection} Several participants (12 out~\num) mentioned that a browser disclosure should make it clear that protecting against website tracking is \textit{not} a security goal of private mode. Five participants argued that Brave has been working on reducing online tracking as a browser feature, and not as a private mode feature.

Further, four participants argued that most browser vendors do not have the incentive to implement a private browsing mode that delivers the level of privacy expected by consumers (see Section~\ref{R3}) -- mainly because most web browsers (\eg, Chrome, Internet Explorer) are owned by companies (\eg, Google, Microsoft) that rely on targeting users with advertisements to generate revenue. Hence, participants explained that disclosures should not use the term ``tracking protection'' to advertise the use of private mode.

\point{Chrome performed better} Many participants (18 out of~\num) perceived the Chrome browser disclosure as \textit{relatively} more informative when compared to the disclosures of Brave and Firefox, as it uses a list of bullet points to describe both private browsing functionality and attackers. In contrast, nine participants reported that the Brave and Firefox disclosures gave them the false sense that private browsing aims to protect against website tracking and ad targeting, increasing their expectations of the protection offered by private mode beyond reality. Also, eight participants mentioned they would use the private mode of Brave and Firefox to perform sensitive browsing activities (before they were given our tutorial), due to the use of the following strong statement by Brave: ``Private tabs \dots always \emph{vanish} when the browser is closed,'' and the use of the shield icon by Firefox. Participants explained that both the statement and the shield are misleading, and do not communicate the actual benefits of private mode. 

Finally, we asked our participants to purpose new disclosure designs to better communicate the benefits and limitations of private mode in different browsers. We discuss the findings in the next section. We also extract a set of design recommendations to help improve the design of disclosures.
\section{Discussion} 
\label{Discussion}

The high-level description of private mode as a ``private browsing tab'' or a ``private browsing window'' is not only vague, but also misleading. Our findings suggest that users' mental models of the term ``private browsing'' influence their understanding and usage of private mode. Incorrect or inappropriate mental models -- partially derived from this term -- could lead users to overestimate the benefits of private mode. For example, some of our participants used private mode to visit webpages not running HTTPS with a valid TLS certificate, incorrectly believing that private mode encrypted Internet traffic. We also found that several participants thought of private mode in connection with end-to-end encrypted communication tools, Tor, and VPN.

Further, only three participants -- who identified as computer security and privacy experts -- correctly explained the primary security goal of private mode. The vast majority of participants incorrectly believed that private mode protected against \emph{any} local attacker, without considering the scenario of a motivated local attacker who could infect a shared machine with a spyware and monitor the user's private browsing activities. 

Therefore, it is critical to communicate the actual protection private mode offers. Although users might learn about private mode from peers and online articles, effective disclosures remain the vendor's most reliable channel to communicate information to users. Hence, drawing from the findings of our study and the browser disclosure designs our participants proposed, we distill the following set of design recommendations that we encourage browser designers to validate, in order to design more effective disclosures related to private mode:

\point{Explain the primary security goal} As most participants pointed out, none of the three browser disclosures they critiqued explained the main security goal of private mode. Although the Google Chrome disclosure says: ``Other people who use this device won't see your activity,'' it does not describe that a malicious user of the device could monitor the private browsing activities of other users through a spyware or a key-logger. Hence, disclosures should clearly explain that private mode only protects against an entity that takes control of the user's machine \textit{after} the user exits private mode.

\point{Explain \emph{where} information about websites visited in private mode is saved} All three browser disclosures have the following statement: ``[Brave; Chrome; Firefox] will not save the following information: your browsing history, ….'' However, several participants argued that this statement is misleading because it implies the information will not be stored by the browser vendor on its servers. Browser designers should consider rewriting the statement to capture the intended meaning: information will not be \emph{locally} stored on the user's device.

\point{Explain \emph{when} information will be deleted} Several participants pointed out that the browser disclosures of both Chrome and Firefox do not explain when information (\eg, browsing history, cookies) about the websites visited in private mode gets deleted. Further, some participants mentioned that although the Brave disclosure says: ``[information] always vanish when the browser is closed,'' it does not clearly communicate the actual functionality of private browsing: information related to a specific private browsing session gets deleted after the user terminates that session. Thus, browser designers should better communicate when private mode-related information will be removed. 

\point{Explain the different types of attackers} Private browsing does not hide activities performed in private mode from motivated local attackers, web attackers, employers, ISPs, browser vendors, and governments (see Section~\ref{PBM}). All three critiqued browser disclosures mention a subset of these attackers. Further, several participants mentioned that disclosures need to clearly describe the entities it can and cannot protect against before explaining the detailed functionality of private mode, as we explain next.

\point{Defer or hide the explanation of functionality} All three disclosures mention different types of files (\eg, browsing history file, cookies, temporary files) that get deleted after the user exits private mode. However, the vast majority of participants did not feel confident explaining what these files meant. Further, several participants preferred that disclosures defer (or hide) the explanation of private browsing functionality until the different types of attackers are described, which none of the critiqued disclosures does.

\point{Avoid using uncertain or misleading words} The Chrome disclosure has the following statement: ``Your activity might still be visible to [the websites you visit, your employer, etc.].'' According to many participants, the use of the word ``might'' could lead users to incorrectly believe that private mode protects against, for example, website tracking. 

Further, the Brave disclosure states the following: ``Private tabs \dots always \emph{vanish} when the browser is closed.'' However, it does not explain \textit{from where} the information gets deleted. The use of the word ``vanish'' led several participants to think that information completely gets removed from local devices and web servers.

\point{Explain the utility of private mode} Most participants did not necessarily use private mode to visit ``embarrassing websites.'' They used the mode to login into an online service using another account, debug/test software, or purchase a surprise gift for a family member or a friend. Hence, some participants suggested that browser disclosures should promote the utility of private mode: what the mode can be used for. 

\point{Use bullet points and bold fonts} In line with prior work, most participants used bullet points in their disclosure designs to explain the functionality and utility of private mode. Our participants also used bold fonts to emphasize important points (mainly, the primary security goal of private mode). 

\point{Notify users when authenticated} We found \textit{all} participants used private mode while being authenticated to online services, incorrectly thinking their search history would get deleted as soon as they exited private mode. Several participants noted they would like to see a mechanism warning them when they start browsing in private mode while being logged into a service.

\point{Rethink the name ``private browsing''} As our findings suggest, the name ``private browsing'' is misleading. Most participants were ``shocked'' and felt ``vulnerable'' upon learning the actual benefits and limitations of private mode. They also suggested different names for private mode, but without a clear winner. Hence, further work should investigate a new name for private mode that would capture its proper usage.

Finally, we encourage browser designers to consider the recommendations we proposed, and design various browser disclosure prototypes. The prototypes can then be validated through designing and conducting future user studies. \textbf{One possible prototype would be to explain the primary security goal of private mode first, followed by a list of bullet points debunking the myths (or misconceptions) that users have about private mode.}
\section{Limitations} \label{LIM} Our study has a number of limitations common to all qualitative research studies. First, the quality of qualitative research mainly depends on the interviewer's individual skills. Therefore, to minimize bias, one researcher, who was trained to conduct interviews and ask questions in an open and neutral way, conducted all~\nn unstructured and~\num semi-structured interviews, as well as all~\nn cognitive interviews (for quiz pre-testing).

Second, some participants’ answers tended to be less detailed. However, the interviewer prompted participants to give full answers to all questions. Further, the interviewer gave participants a~10-minute break between the second (interviews) and third (participatory design) parts of the study, to reduce interviewee fatigue and inattention~\cite{DS} (see Section~\ref{PILOT}).

Third, as with all qualitative studies, our work is limited by the size and diversity of our sample. Following recommendations from prior work to interview between~12 and~\num participants~\cite{L1}, we interviewed participants until new themes stopped emerging (total:~\num participants). We also recruited a demographically-diverse sample of participants in order to increase the likelihood that relevant findings have been mentioned by at least one participant.
\section{Conclusion} \label{Conc} In this work, we investigated why most users misunderstand the benefits and limitations of private mode. We did so by designing and conducting a three-part study. We recruited~\num demographically-diverse participants, who used or had used in the past private mode, for the second and third parts of the study. We first performed a usability inspection of private mode using both cognitive walkthrough and heuristic evaluation. We then conducted a qualitative user study to explore users’ mental models of private mode and its security goals. We finally performed a participatory design study to investigate why existing browser disclosures misinform users of the actual protection offered by private mode. 


\bibliographystyle{IEEEtran}
\raggedright
\bibliography{MyBib}

\vspace{0.5ex}
\appendix
\subsection{Usability Inspection: Hybrid Approach} \label{a}
We here describe the hybrid approach we used to inspect the user interface of private mode in web browsers:
\begin{enumerate}
    \item Provide a detailed description of the user interface.
    \item Define the users and their goals.
    \item Define the tasks the users would attempt (\eg, accessing a web page in private mode).
    \item Break each task into a sequence of sub-tasks or actions (\eg, selecting the “New Private Window” option).
    \item Walk through each task workflow step-by-step through the lens of the users (\eg, what they would look for, what paths they would take, what terms they would use).
    \item For each action, look for and identify usability problems based on a set of heuristics.
    \item Specify where the usability problem is in the user interface, how severe it is, and possible design fixes.
\end{enumerate}

\vspace*{0.5ex}
\subsection{Unstructured Interview Participant Demographics} \label{b}
\begin{table}[htbp]
\caption{Unstructured Interview Participant Demographics}
    \scriptsize
    \centering
    \setlength{\arrayrulewidth}{.2em}
    \setlength{\tabcolsep}{8pt}
    \begin{tabular}{lllll}
    \toprule
    \textbf{Gender}&\textbf{Age}&\textbf{Race}&\textbf{Education}&\textbf{Employment}\\ \midrule
    Male& 18--24& Asian& Some college& Student\\
    Male& 35--44& Hispanic& B.Sc.& Employed\\
    Female& 25--34& White& M.Sc.& Student\\
    Male& 18--24& White& B.Sc.& Student\\
    Female& 55--64& Black& B.A.& Retired\\
    \bottomrule
    \end{tabular}
\label{T1}
\end{table}

\vspace*{0.5ex}
\subsection{Pilot Study: Cognitive Interview Participant Demographics} \label{c}
\begin{table}[htbp]
\caption{Cognitive Interview Participant Demographics}
    \scriptsize
    \centering
    \setlength{\arrayrulewidth}{.2em}
    \begin{tabular}{lllll}
        \toprule
        \textbf{Gender}&\textbf{Age}&\textbf{Race}&\textbf{Education}&\textbf{Employment}\\ \midrule
        Male& 18--24& Black& B.Sc.& Student\\
        Male& 35--44& Asian& M.Sc.& Employed\\
        Female& 18--24& White& B.Sc.& Student\\
        Male& 55--64& White& Some college& Retired\\
        Female& 45--54& Hispanic& Some college& Employed\\
        \bottomrule
    \end{tabular}
    \vspace{-1em}
    \label{T2}
\end{table} 

\vspace*{0.5ex}
\newpage
\onecolumn
\subsection{Selected Participant Mental Models of ``Private Browsing’’} \label{e}
\begin{figure*}[!htb]
\centering
\begin{minipage}{0.31\textwidth}
\centering
\includegraphics[trim={0 0cm 0 0}, width=0.9\linewidth, clip]{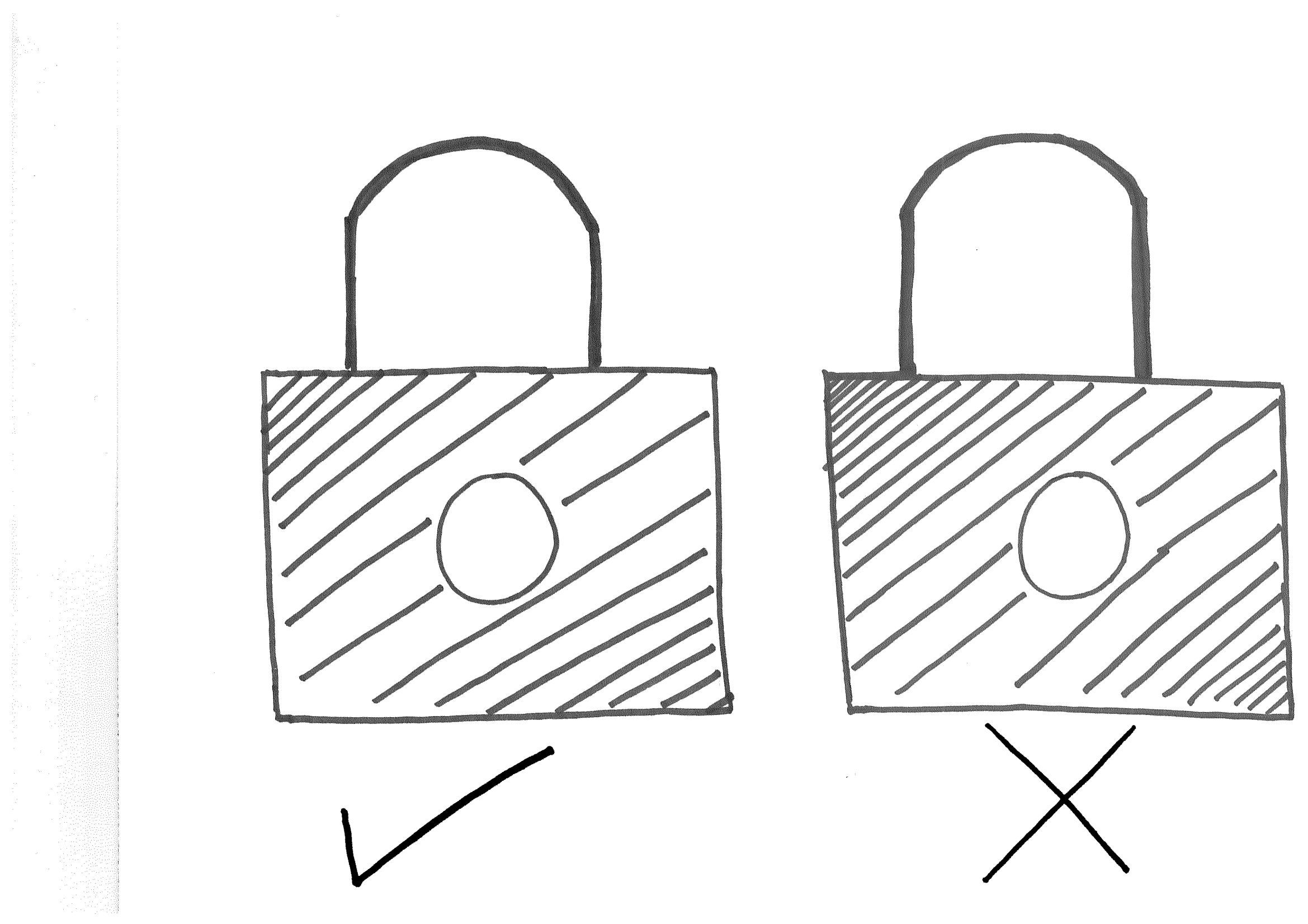}
\caption{Secure/encrypted browser connections.}
\label{Fig7}
\end{minipage}
\begin{minipage}{0.31\textwidth}
\centering
\includegraphics[width=0.9\linewidth]{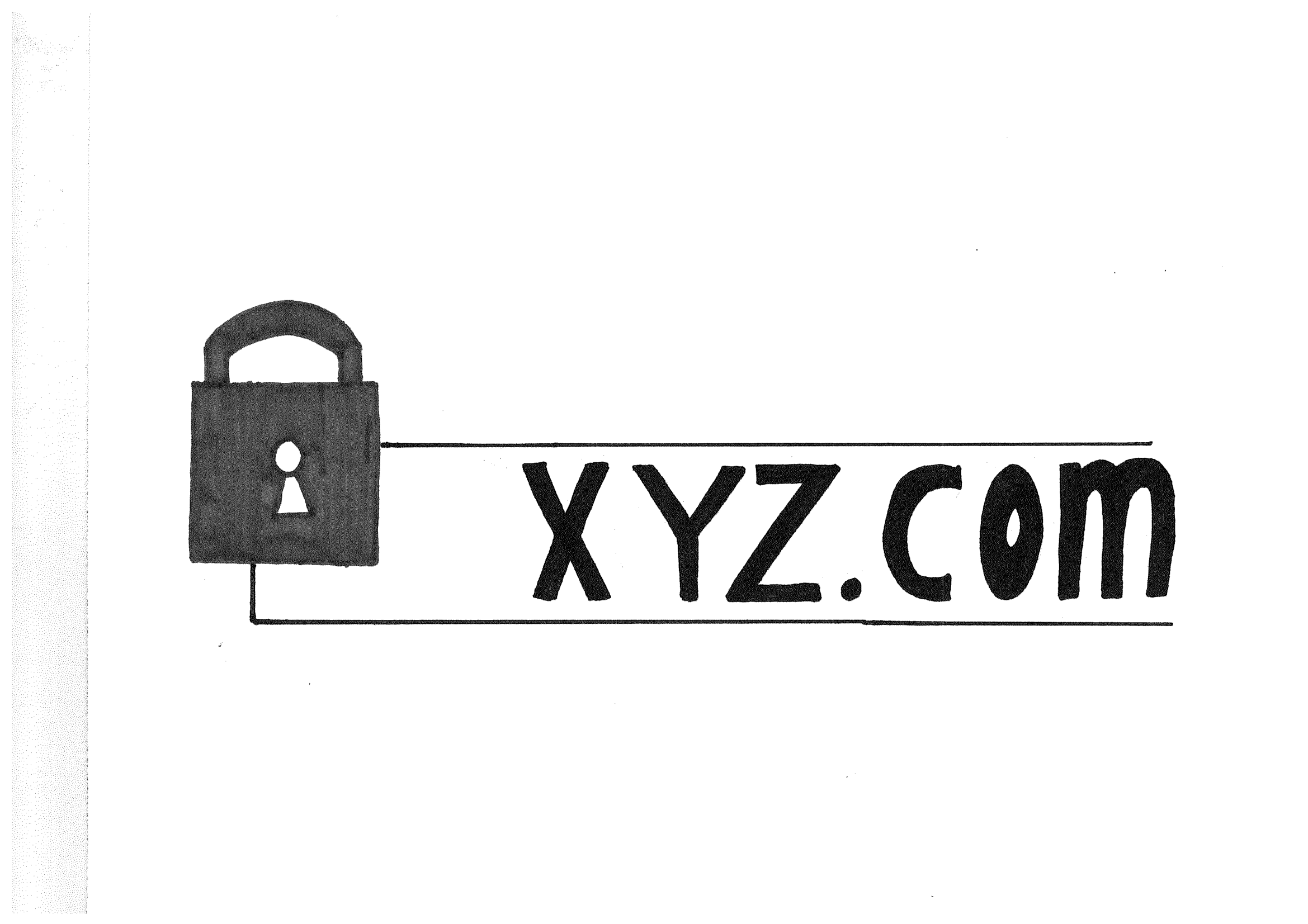}
\caption{Secure/encrypted browser connections.}
\label{Fig8}
\end{minipage}
\begin{minipage}{0.31\textwidth}
\centering
\includegraphics[trim={0 0cm 0 0}, width=0.9\linewidth, clip]{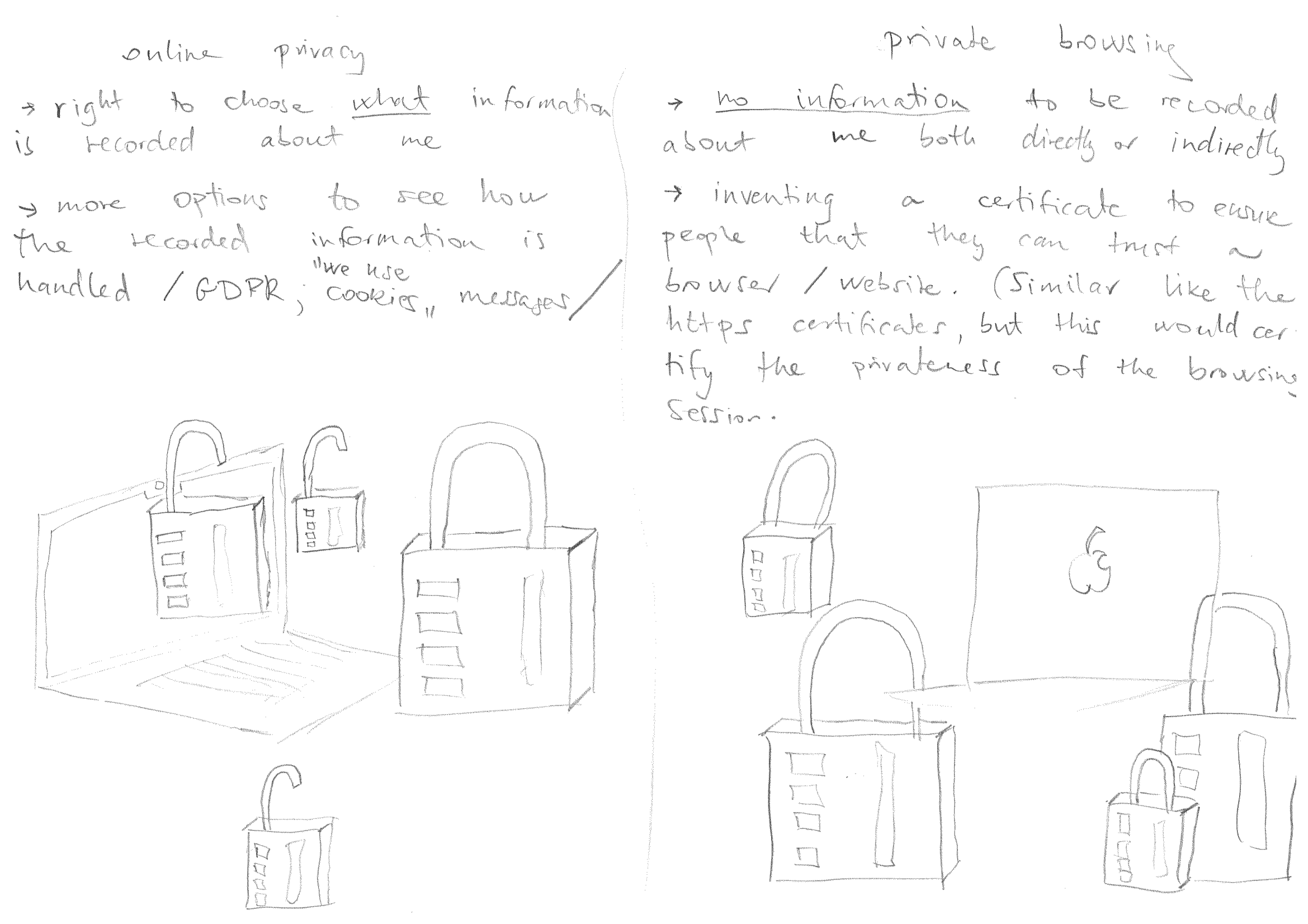}
\caption{Secure/encrypted browser connections.}
\label{Fig9}
\end{minipage}
\centering
\end{figure*}

\begin{figure*}[!htb]
\centering
\begin{minipage}{0.31\textwidth}
\centering
\includegraphics[trim={0 0cm 0 0}, width=0.9\linewidth, clip]{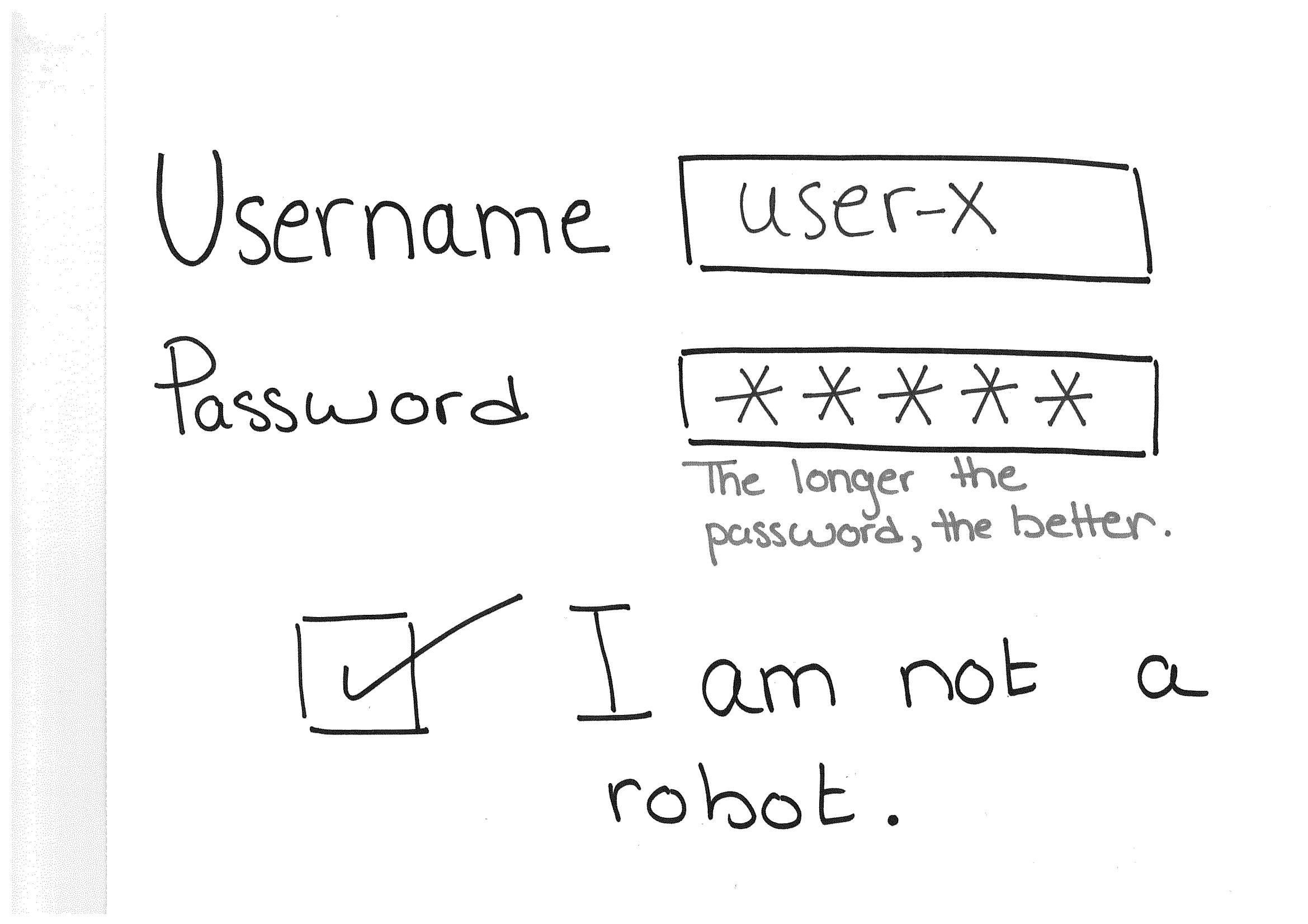}
\caption{One-factor authentication.}
\label{Fig7}
\end{minipage}
\begin{minipage}{0.31\textwidth}
\centering
\includegraphics[width=0.9\linewidth]{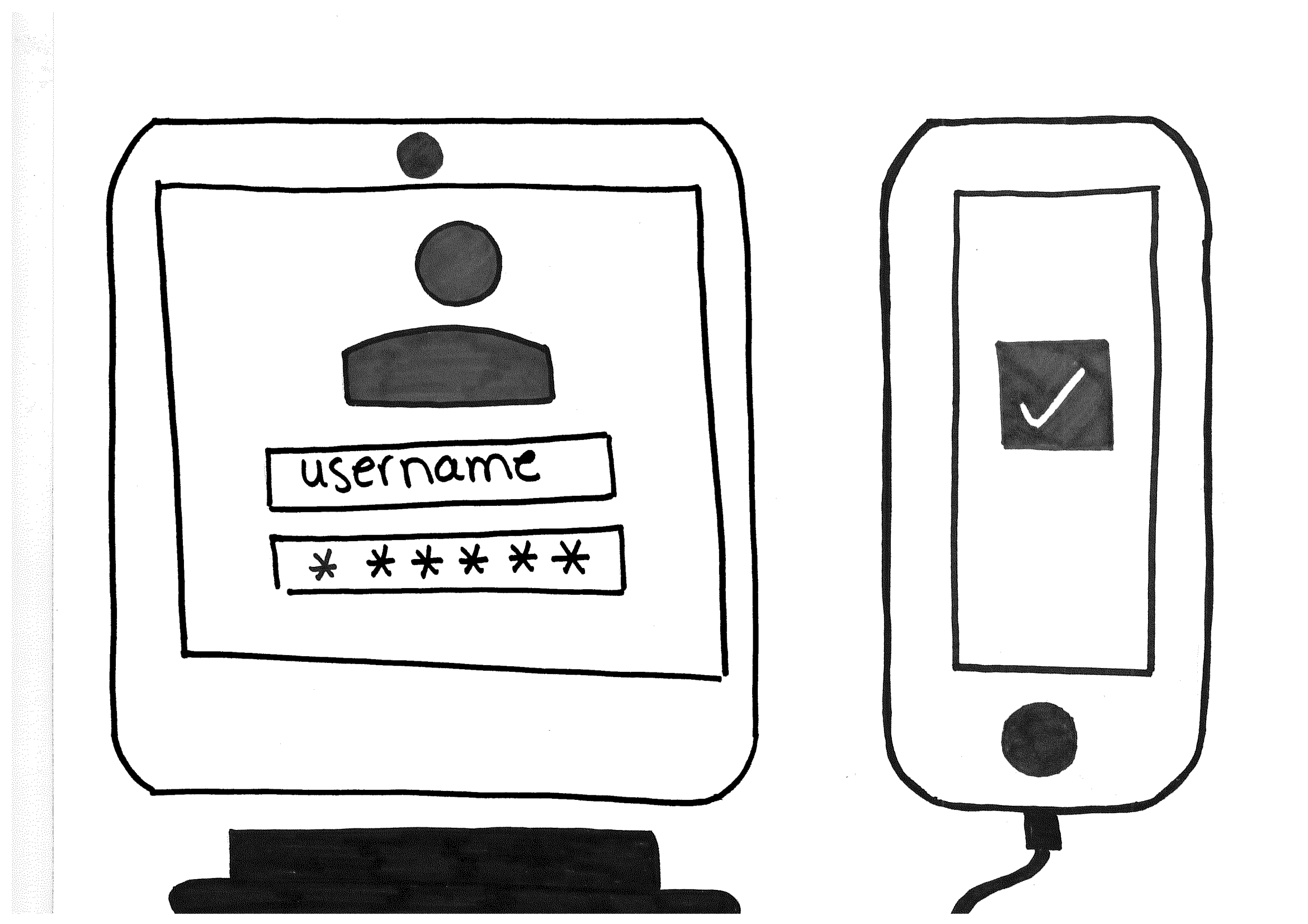}
\caption{Two-factor authentication.}
\label{Fig8}
\end{minipage}
\begin{minipage}{0.31\textwidth}
\centering
\includegraphics[trim={0 0cm 0 0}, width=0.9\linewidth, clip]{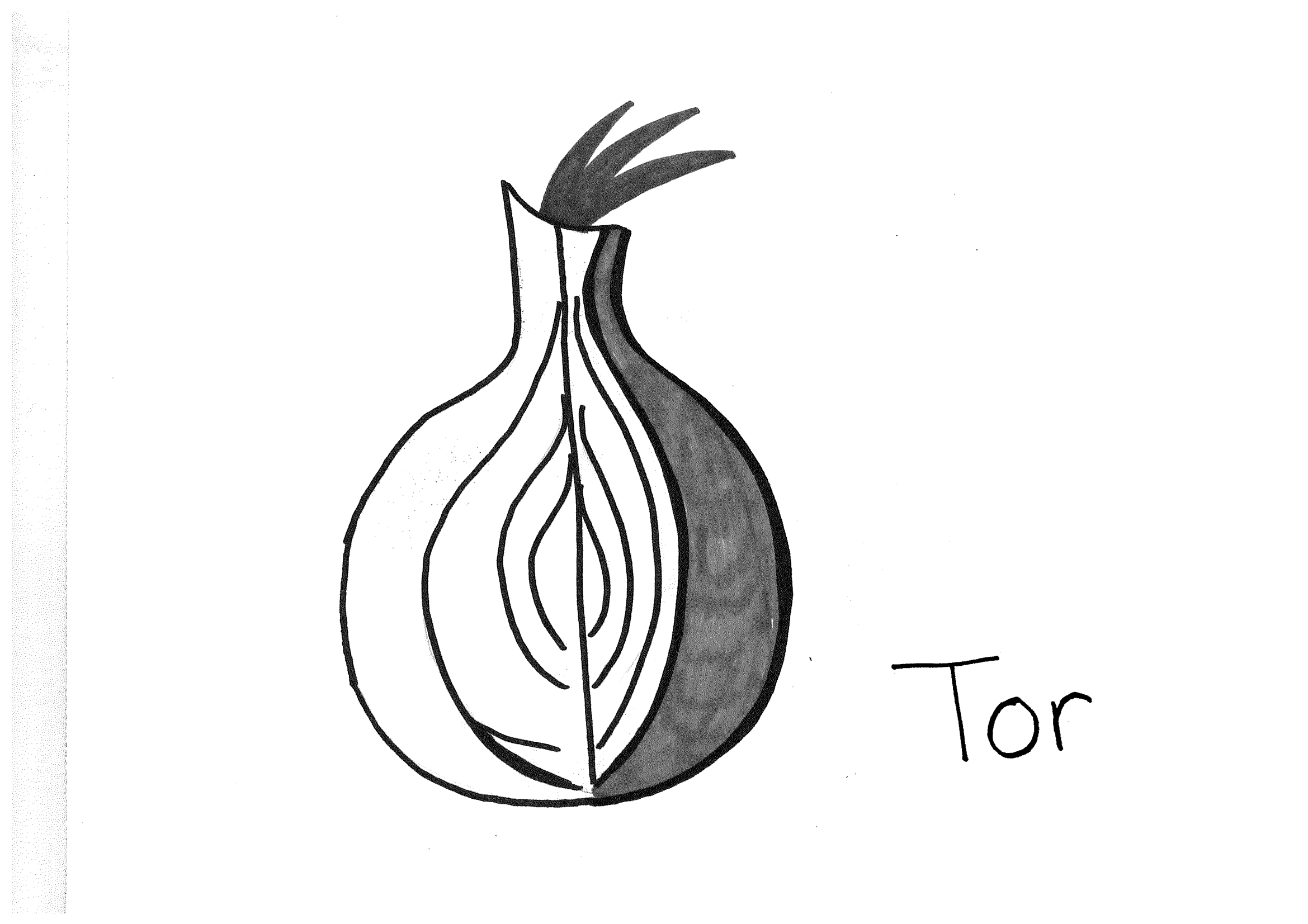}
\caption{Anonymous browsing (using Tor).}
\label{Fig7}
\end{minipage}
\centering
\end{figure*}

\begin{figure*}[!htb]
\centering
\begin{minipage}{0.31\textwidth}
\centering
\includegraphics[trim={0 0cm 0 0}, width=0.9\linewidth, clip]{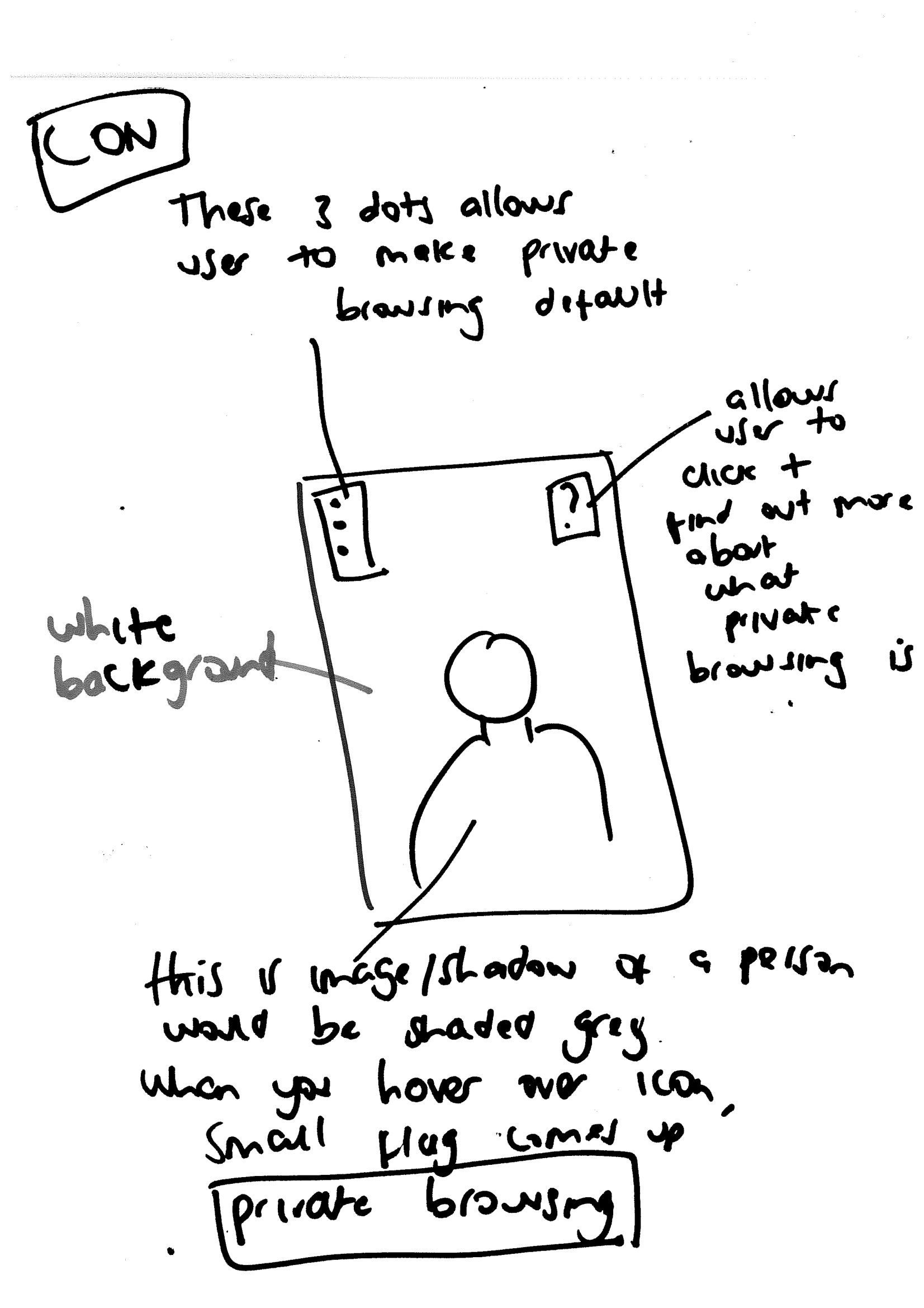}
\caption{Private mode.}
\label{Fig7}
\end{minipage}
\begin{minipage}{0.31\textwidth}
\centering
\includegraphics[width=0.9\linewidth]{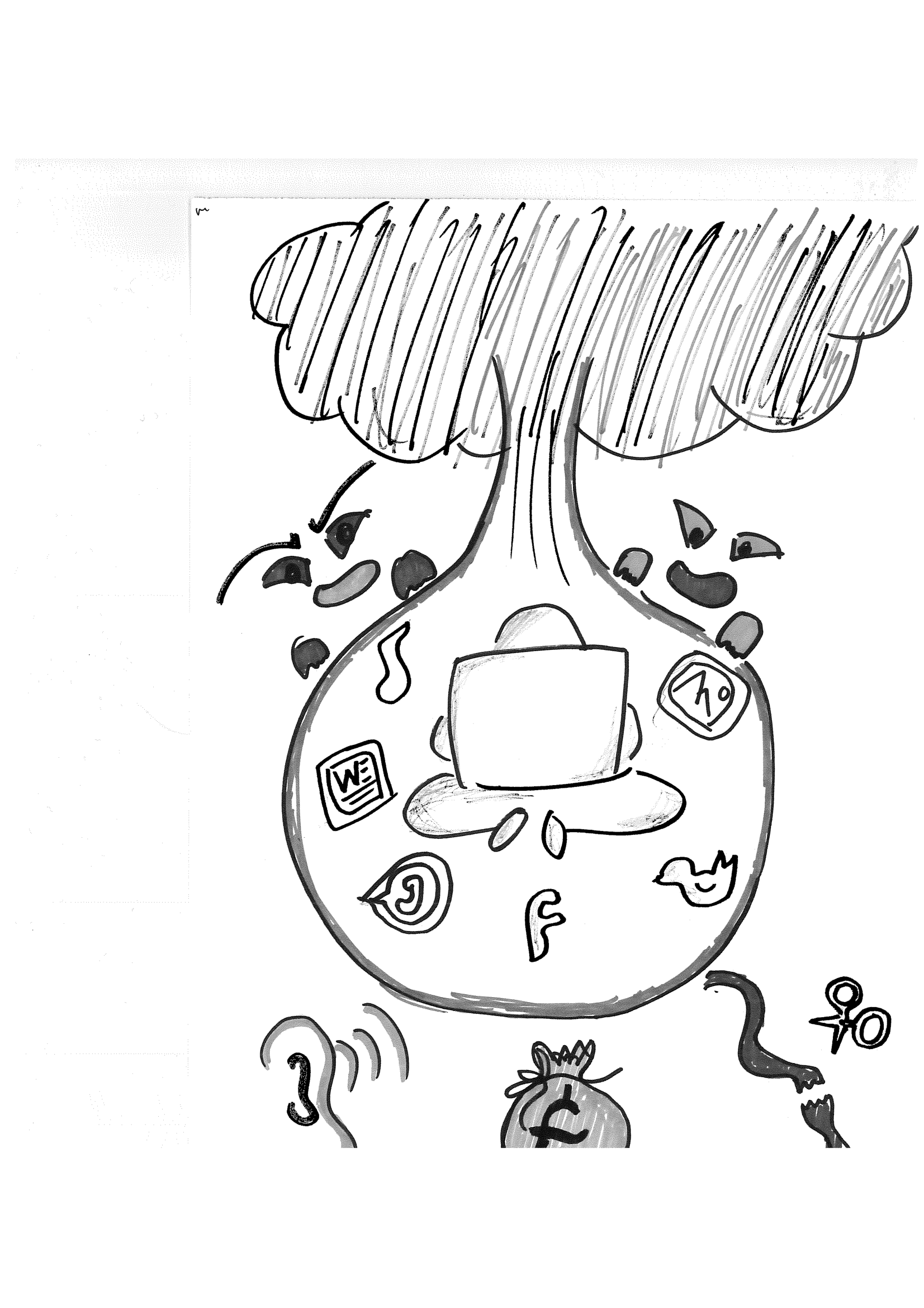}
\caption{Complete online privacy.}
\label{Fig8}
\end{minipage}
\begin{minipage}{0.31\textwidth}
\centering
\includegraphics[trim={0 0cm 0 0}, width=0.9\linewidth, clip]{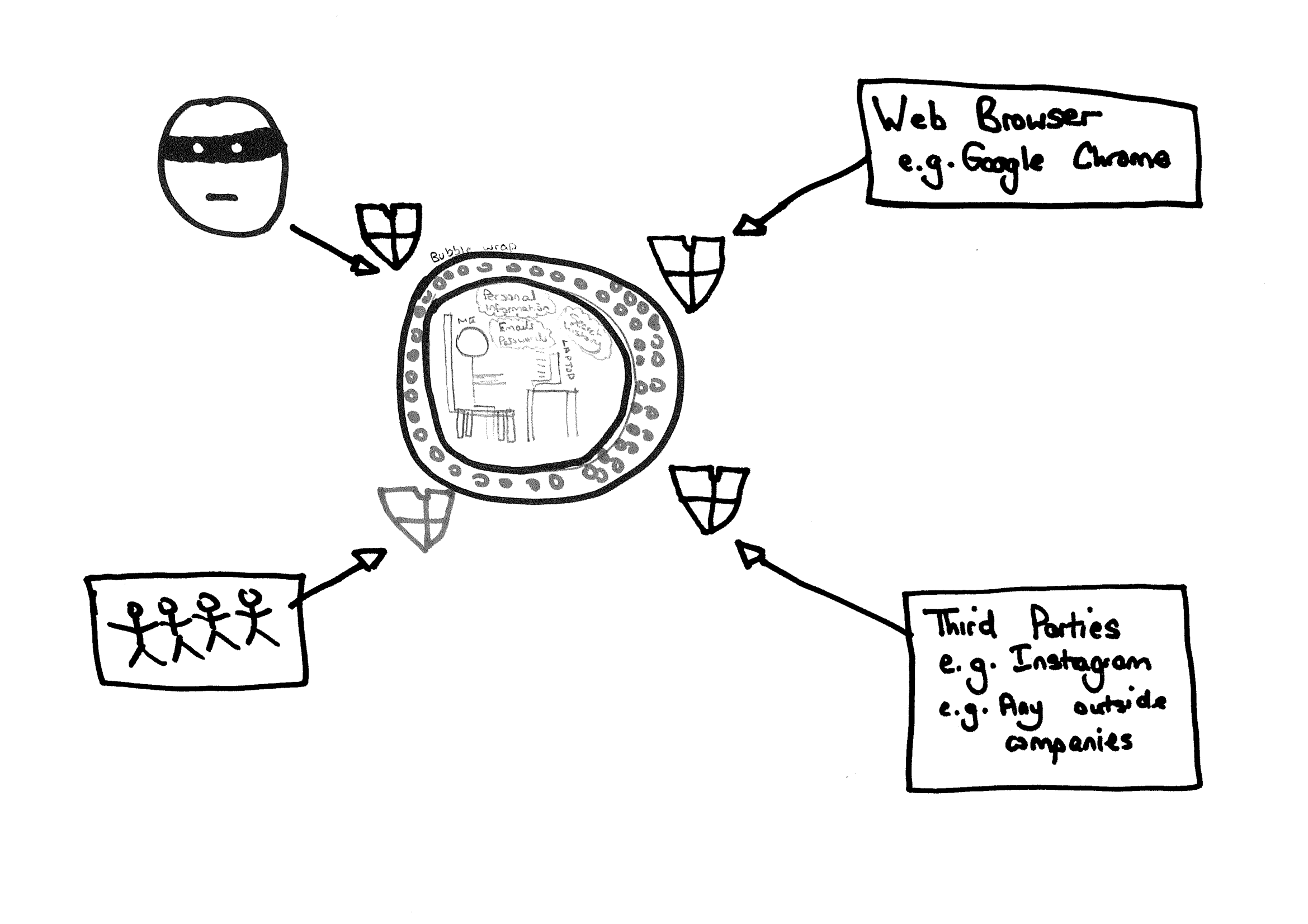}
\caption{Complete online privacy.}
\label{Fig7}
\end{minipage}
\centering
\end{figure*}

\vspace*{0.5ex}
\newpage
\onecolumn
\begin{sidewaystable}[h]
\subsection{Studies of Private Mode} \label{e}
\caption{A Detailed Overview of User Studies of Private Browsing Mode}
    \centering
    \setlength{\arrayrulewidth}{.2em}
    \resizebox{\textwidth}{!}{
    \begin{tabular}{lL{0.5\textwidth}L{0.5\textwidth}L{0.5\textwidth}L{0.5\textwidth}L{0.5\textwidth}}
    \toprule
    &\textbf{Study} &\textbf{Research Questions} &\textbf{Methodology} &\textbf{Key Findings} &\textbf{Recommendations}\\ \midrule
    \addlinespace[0.5cm]
    \textbf{1}&\textbf{An Analysis of Private Browsing Modes in Modern Browsers (USENIX Security, 2010)~\cite{PB}} &\begin{itemize}
        \item Are people aware of private browsing?
        \item How often do people use private browsing?
        \item Do users of a specific web browser use private mode more frequently than, as frequently as, or less frequently than users of another web browser? 
        \item What do people use private browsing for?
    \end{itemize} &\begin{itemize}
        \item Study type:~\textbf{A measurement study (quantitative).}
        \item Aggarwal \etal performed the first measurement study to monitor people's private browsing usage in four browsers (Firefox, Google Chrome, Internet Explorer, and Safari) on three different types of websites (adult, online shopping, and news).
        \item The measurement software detected if a website was visited in public or private mode.
        \item They ran three simultaneous one-day campaigns targeting adult, gift shopping, and news websites.
        \item They collected~155,226 impressions.
    \end{itemize} &\begin{itemize}
        \item Participants often used private browsing to visit adult websites, and not online shopping or news websites.
        \item Firefox~3.6 and Safari~4.0 had high rates of private browsing usage, compared to Google Chrome~4.0 and Internet Explorer~8.0. Aggarwal \etal argue web browsers that do not have a visual user interface element that clearly indicates a user is currently browsing in private mode lead users to open a private tab or window and forget to close it, explaining the high rates of private browsing usage in Firefox~3.6 and Safari~4.0. 
    \end{itemize} &\begin{itemize}
        \item No recommendations were provided.
    \end{itemize}\\
    
    \addlinespace[0.5cm] \textbf{2} &\textbf{Understanding Private Browsing (a study by Mozilla, 2010)~\cite{PB0}} &\begin{itemize}
        \item At what time of the day do people (who are aware of private browsing) use private mode?
        \item How long do people stay in a private browsing session?
    \end{itemize} &\begin{itemize}
        \item Study type:~\textbf{A measurement study (quantitative).}
        \item Mozilla conducted a test pilot study to record the time Firefox~3.5 users activated private browsing, as well as the time they deactivated it.
        \item Test Pilot was developed as an opt-in service for Firefox Beta users.
        \item The study did not indicate the number of Beta users who opted-in. 
    \end{itemize} &\begin{itemize}
        \item Participants likely browsed in private mode during lunchtime (between 11:00 am and 2:00 pm) and after they had returned from school or work (around 5:00 pm).
        \item Participants usually stayed in a private browsing session for about~10 minutes.
        \item The duration of a private browsing session did not considerably fluctuate throughout the day.
    \end{itemize} &\begin{itemize}
        \item No recommendations were provided.
    \end{itemize}\\
    
    \addlinespace[0.5cm]
    \textbf{3} &\textbf{Private Browsing: An Inquiry on Usability and Privacy Protection (WPES, 2014)~\cite{PB1}} &\begin{itemize}
        \item Are people aware of private browsing?
        \item What do people use private browsing for?
        \item At what time of the day do people browse in private mode?
        \item How do people perceive the benefits and drawbacks of private browsing?
    \end{itemize} &\begin{itemize}
        \item Study type:~\textbf{A survey (quantitative).}
        \item Gao \etal conducted a survey of~200 US respondents (via MTurk).
    \end{itemize} &\begin{itemize}
        \item About one-third of respondents were not aware of private browsing.
        \item Respondents who had used private browsing mentioned using it for visiting adult websites, online shopping, and avoiding website tracking.
        \item Respondents reported using private browsing during work, or at night (after they had returned from work).
        \item Some respondents who were aware of, and/or had used, private browsing incorrectly believed that private mode hid their private browsing activities from visited websites.
    \end{itemize} &\begin{itemize}
        \item The name ``private browsing'' should be rethought.
        \item Browser disclosures related to private browsing should be redesigned to better inform users of the benefits and limitations of private browsing.
    \end{itemize}\\
    
    \addlinespace[0.5cm] \textbf{4} &\textbf{A Study on Private Browsing: Consumer Usage, Knowledge, and Thoughts (a study by DuckDuckGo, 2017)~\cite{PB2}} &\begin{itemize}
        \item Are people aware of private browsing?
        \item How do people use private browsing?
        \item What do people use private browsing for?
        \item How do people perceive the benefits and drawbacks of private browsing?
        \item How do people react to private browsing knowledge?
    \end{itemize} &\begin{itemize}
        \item Study type:~\textbf{A survey (quantitative).}
        \item DuckDuckGo conducted a survey of~5,710 US respondents (via SurveyMonkey).
    \end{itemize}& \begin{itemize}
        \item About one-third of respondents had not heard of private browsing.
        \item About one-half of respondents had used private browsing at least once.
        \item Respondents used private browsing on both desktop and mobile phone.
        \item Most respondents used private browsing to visit ``embarrassing websites.''
        \item About three-quarters of respondents were not able to correctly identify the benefits and limitations of private browsing. Further, two-thirds overestimated the benefits of private browsing.
        \item Some respondents incorrectly thought that private browsing prevented visited websites from tracking them, as well as search engines from knowing their searches.
        \item About two-thirds of respondents felt ``surprised'' or ``vulnerable'' upon learning about the actual protections of private browsing.
    \end{itemize} &\begin{itemize}
        \item No recommendations were provided.
    \end{itemize}\\
    
    \addlinespace[0.5cm] \textbf{5} &\textbf{Understanding Why People Use Private Browsing (a study By Elie Bursztein, 2017)~\cite{PB3}} &\begin{itemize}
        \item Are people aware of private browsing, and do they use it?
        \item What do people use private browsing for?
        \item Where do people use private browsing?
        \item Who do people hide from when using private browsing?
    \end{itemize} &\begin{itemize}
        \item Study type:~\textbf{A survey (quantitative).}
        \item Bursztein ran a survey of~200 US respondents (via Google Consumer Surveys).
    \end{itemize} &\begin{itemize}
        \item About one-third of respondents did not know what private browsing is.
        \item Only one-fifth reported using private browsing.
        \item One-half of respondents preferred not to disclose what they used private browsing for. One-fifth reported using it for online shopping.
        \item Respondents reported using private browsing to hide their browsing activities from people sharing their computer, their ISP, and visited websites.
    \end{itemize} &\begin{itemize}
        \item Surveys are~\textbf{not} the best research method to elicit users’ private browsing habits due to the “embarrassing factor.''
        \item The computer security and privacy community should raise awareness of the benefits and limitations of private browsing, to enable users to make informed decisions.
    \end{itemize}\\
    
    \addlinespace[0.5cm] \textbf{6} &\textbf{Your Secrets Are Safe: How Browsers' Explanations Impact Misconceptions About Private Browsing Mode (WWW, 2018)~\cite{PB4}} &\begin{itemize}
        \item Prior work has shown that users have several misconceptions about private browsing, but do browser disclosures (related to private browsing) contribute to these misconceptions?
    \end{itemize} &\begin{itemize}
        \item Study type:~\textbf{A survey (quantitative).}
        \item Wu \etal conducted a survey of~460 US respondents (recruited via MTurk).
        \item Respondents were assigned one of~13 disclosures of different web browsers.
        \item Based on the disclosure they saw, respondents were asked to answer a set of questions about what would happen to different items (\eg browsing history entries, cookies, downloaded files) when browsing in public and private modes.
    \end{itemize} &\begin{itemize}
        \item The Google Chrome desktop disclosure led respondents to answer more questions correctly. However, \emph{all} tested browser disclosures failed to correct users' misconceptions about private browsing.
    \end{itemize} &\begin{itemize}
        \item Browser disclosures should be redesigned to better communicate the actual protections of private browsing to users.
    \end{itemize}\\
    
    \addlinespace[0.5cm] \textbf{7} &\textbf{Away From Prying Eyes: Analyzing Usage and Understanding of Private Browsing (SOUPS, 2018)~\cite{PB5}} &\begin{itemize}
        \item How do people use private browsing?
        \item What do people use private browsing for?
        \item Are people at risk when using private browsing?
    \end{itemize} &\begin{itemize}
        \item Study type:~\textbf{A measurement study and a survey (quantitative)}
        \item Habib \etal conducted a user study of~460 US participants who used the Security Behaviour Observatory (SBO), a panel that actively collects data related to security and privacy behaviour of users.
        \item They distributed a follow-up survey (via SBO and MTurk), to explore discrepancies, if any, between observed and self-reported private browsing behaviour.
    \end{itemize} &\begin{itemize}
        \item Only~4\% of SBO participants used private browsing.
        \item The most common private browsing activities (\eg visiting adult websites, online shopping, logging into an online service) were the same across both observed and self-reported data.
        \item Many participants overestimated the benefits of private browsing.
    \end{itemize} &\begin{itemize}
        \item Browser disclosures should be redesigned.
    \end{itemize}\\
    
    \addlinespace[0.5cm] \textbf{8} &\textbf{Evaluating the End-User Experience of Private Browsing Mode (our study)} &\begin{itemize}
        \item Does private mode in different web browsers suffer from poor usability that hampers the widespread adoption and use of private browsing?
        \item How do people perceive the term “private browsing?''
        \item What are people’s mental models of private browsing (as a privacy-enhancing technology) and its security goals?
        \item How do people perceive those who use private browsing? Do people perceive the routine use of private browsing as “paranoid” or “unnecessary?”
        \item How do people's mental models and perceptions influence their usage of private browsing?
        \item Why do existing browser disclosures (related to private browsing) misinform people of the benefits and limitations of private browsing?
        \item How can the design of browser disclosures be improved?
    \end{itemize} &\begin{itemize}
        \item Study type:~\textbf{a usability inspection + a qualitative study.}
        \item We conducted a three-part study: (1) a usability inspection of private mode in different web browsers; (2) a qualitative, interview-based study; (3) a participatory design study.
    \end{itemize} &\begin{itemize}
        \item The user interface of private mode violates several design principles and heuristics.
        \item Participants’ conceptual understanding of the term “private browsing” influenced their understanding and usage of private mode in real life. 
        \item Almost all participants did not understand the primary security goal of private browsing. 
        \item Some participants perceived those who used private mode as ``paranoid,'' ``having something to hide,'' or ``up to no good.''
        \item Participants critiqued existing browser disclosures and designed new ones.
    \end{itemize} &\begin{itemize}
        \item The key user-related challenge for private browsing is not adoption, but appropriate use.
        \item We distilled a set of design recommendations to help browser designers design better and more effective browser disclosures.
    \end{itemize}\\
    \bottomrule
    \end{tabular}}
    \label{TRelatedWork}
\end{sidewaystable}

\end{document}